\documentclass[prb,floatfix,amsmath,preprint]{revtex4}
\usepackage{epsfig}
\usepackage{graphicx}
\usepackage{dcolumn}
\usepackage{bm}
\begin{document}
\title{Micro-plasticity and intermittent dislocation activity in a simplified micro structural model}
\author{P. M. Derlet}
\email{Peter.Derlet@psi.ch} 
\affiliation{Condensed Matter Theory Group, Paul Scherrer Institut, CH-5232 Villigen PSI, Switzerland}
\author{R. Maa{\ss}}
\email{maass@caltech.edu}
\altaffiliation{Financially supported by the Alexander von Humboldt Foundation.}
\affiliation{California Institute of Technology, Division of Engineering and Applied Sciences, 1200 E California Blvd, Pasadena, CA 91125-810
0, USA}
\date{\today}

\begin{abstract}
Here we present a model to study the micro-plastic regime of a stress-strain curve. In this model an explicit dislocation population represents the mobile dislocation content and an internal shear-stress field represents a mean-field description of the immobile dislocation content. The mobile dislocations are constrained to a simple dipolar mat geometry and modelled via a dislocation dynamics algorithm, whilst the shear-stress field is chosen to be a sinusoidal function of distance along the mat direction. The sinusoidal function, defined by a periodic length and a shear-stress amplitude, is interpreted to represent a pre-existing micro-structure. These model parameters, along with the mobile dislocation density, are found to admit a diversity of micro-plastic behaviour involving intermittent plasticity in the form of a scale-free avalanche phenomenon, with an exponent and scaling-collapse for the strain burst magnitude distribution that is in agreement with mean-field theory and similar to that seen in experiment and more complex dislocation dynamics simulations.
\end{abstract}

\maketitle

\section{Introduction}

In 1964, using state-of-the-art torsion experiments, Tinder and co-workers were able to achieve a strain resolution of $10^{-8}$ for sub mm-sized samples to study the micro-plastic regime of highly pure poly-crystal Cu samples \cite{Tinder1964}, followed by tests on Zn single crystals some years later \cite{Tinder1973}. Their plastic-strain versus stress curves contained plateaus of stress which were attributed to the occurrence of discrete dislocation glide activity over a length scale comparable to the dislocation spacing of an assumed three-dimensional network. Indeed the authors write in ref.~\cite{Tinder1964}, ``The results suggest that an important fraction of the total strain, in the initial stages of deformation, involved motion of a few favourably situated dislocation segments through distances large enough to form new interactions with other elements of the three dimensional network. If this were so, then most elements of the network must have been relatively immobile, making little or no contribution to the strain.'' That most dislocations remain immobile remains a contemporary viewpoint~\cite{Neuhaeuser1983}.

Another more recently pursued route is to probe discrete dislocation activity via the stress-strain curve of micron sized focused ion beam (FIB) milled single crystals. Here, advantage is taken of a nano-indentation platform equipped with a flat punch tip to compress micro-/nano-crystals~\cite{Uchic2004,Greer2005,Dimiduk2006,Dimiduk2010}. In addition to the sub-nanometer displacement resolution of the system, which notably has a lower strain resolution than the above mentioned torsion experiments, the sample size is decreased to the micron range and below. As a result, the strain associated with the discrete dislocation activity is increased to an easily detectable magnitude. Whilst this more contemporary work has been primarily motivated by the ``smaller-is-stronger'' size-effect paradigm \cite{Uchic2004,Greer2005,Volkert2006,Maass2011,Maass2012}, an extensive analysis of the statistics of the discrete dislocation activity has revealed power-law behavior in the distribution of strain-burst magnitudes giving an exponent of $\approx1.6-2.2$~\cite{Dimiduk2006,Dimiduk2010,Zaiser2008}. Similar exponents can also be found in bulk samples via detailed analysis of load displacement signals, where structural evolution is also seen to occur~\cite{Fressengeas2009,Mudrock2011}. The very recent work of Dahmen and co-workers~\cite{Dahmen2009,Friedman2012} suggest that the variation in literature exponent values could be due to the different stress intervals used to bin the strain burst magnitude data, where a total integration over the stress interval to material value should, within mean field theory, give an exponent equal to precisely two. Such exponents are indicative of crackling~\cite{Sethna2001} or Barkhausen noise, and more generally of avalanche phenomena, indicating that dislocation mediated plastic deformation belongs to a universality class that encompasses many natural phenomenon over a variety of different length and timescales. Indeed, similar power-law exponents can also be found for metallic glasses~\cite{Sun2010} in which the underlying plastic deformation is fundamentally different to crystalline metals. 

Another class of experiments revealing the intermittent nature of dislocation dynamics is the acoustic emission monitoring of ice~\cite{Miguel2001,Weiss2003}. Such experiments measure the acoustic energy released by intermittent dislocation activity during constant stress deformation (in the tertiary creep regime). Indeed, via single sensor acoustic emission signals, such dislocation activity can be well characterised in time revealing power-law behaviour with exponents of 1.6 to 1.8, which is very similar to that seen via the micro-compression stress-strain curves~\cite{Dimiduk2006,Dimiduk2010,Zaiser2008}. Furthermore, via multiple sensor monitoring, time and space clustering of dislocation avalanches could be observed~\cite{Weiss2003}. It was found that avalanche epi-centres were correlated in space according to a non-integer power-law exponent indicating scale-free clustering and at short enough times such clusters were correlated in time indicating collective activity.

That such scale-invariant dislocation activity occurs, is a signature of an underlying complex dislocation based micro-structure. An entity whose properties and evolution under an applied stress play a central role in the more general subjects of material strength and strain hardening~\cite{DisInSolids}. Due to the complex dynamics and evolution of the dislocation structure, computer simulation based approaches have helped greatly over the past decades to clarify the underlying dislocation based mechanisms responsible for such structural evolution. One such method is the so-called dislocation dynamics (DD) approach. Early work involved two dimensional arrays of straight edge dislocations interacting via elasticity using single- and multi-slip geometries~\cite{Amodeo1990,Devincre1993,Fournet1996}. These works demonstrate that under an external stress, dislocation patterning emerges analogous to what is seen in static transmission electron microscopy experiments. Other and more recent works have developed these numerical techniques in terms of efficiency~\cite{Bako2006}, strain boundary-conditions and obstacle/composite geometries \cite{derGiessen1995} to study more complex patterning such as the emergence of granular cell structures and the study of grain boundary network evolution \cite{Ispanovity2011a}, strain hardening, and material strength in both bulk and confined volume systems \cite{Ispanovity2011b,Deshpande2005}. Analogous methodologies and investigations have also been carried out in three dimensions \cite{Weygand2002,CaiBook,Csikor2007,Rao2008,Devincre2009}.

Intermittent dislocation activity, in the form of dislocation avalanches, has also been studied using the DD simulation technique~\cite{Miguel2001,Ispanovity2011b,Csikor2007}. Such simulations have produced power-law exponents of the distribution of strain burst magnitudes similar to that seen in experiments, indicating some degree of scale-free behaviour, where dislocations are arranged in meta-stable cell and/or wall structures, with only a minor fraction of the dislocation population moving intermittently, thereby creating discrete strain jumps. The so-called self organised criticality (SOC) view of the dislocation network state offers one theoretical platform for the understanding of the observed universality~\cite{Zaiser2006}, in which the material system organises itself into a configuration that is critical, resulting in scale-free behaviour upon transiting to a new realisation of the critical state. Originally developed to describe sand-pile dynamics~\cite{Bak1987}, the approach is somewhat at odds with the historical viewpoint that dislocation structure evolution is primarily driven by equilibrium driving forces such as that embodied in Low Energy Structures (LES) theory~\cite{KuhlmannWilsdorf2002} and in a wide range of strain hardening theories. The finding that the occurrence of avalanche phenomenon is insensitive to the nature of the forming immobile dislocation network, the slip geometry, the deformation mode, and the details of the dislocation dynamics and spatial dimension is a central hallmark of SOC, which is robust against the details of the underlying physical model. 

This motivates investigating plastic flow with simpler models that explicitly do not take into account the fine details of individually interacting sessile and mobile dislocations, naturally shifting the focus of plastic flow from complex dislocation structural evolution to the interaction of a minute mobile dislocation population with a simplified description of the sessile dislocation population. Such an approach is analogous to the study of dislocations in the presence of pinning potentials \cite{Moretti2004,Leoni2009} and more generally to coarse grained models of plasticity that study the depinning transition (see ref.~\cite{Zaiser2006} and references therein, and also refs.~\cite{Dahmen2009,Friedman2012}). Dislocation field theories are also well able to study intermittency, however unlike dislocation dynamics based methods which are able to simulate only small plastic strains, these models are able to incorporate dislocation transport over non-neglible distances and therefore able to simulate significant structural evolution~\cite{Fressengeas2009}.

In this work, a simple model is therefore proposed in which a dipolar dislocation mat~\cite{Hansen1986} is embedded into an internal static sinusoidal stress field defined by a wave-length and a shear-stress amplitude. The explicit dislocation population is modelled via a standard DD algorithm. It is found that, similar to very complex and detailed 3D-DD simulations, the resulting stress-strain curves evolve in a discrete manner that reflects an underlying intermittent plasticity originating from irreversible changes in dislocation configuration, and that the distribution of the corresponding strain-burst magnitudes reveal both extremal value statistics and scale-free avalanche behaviour. Thus, although the complex details of microscopic dislocation mechanisms and structure are omitted, the simple model is still able to capture the fundamental properties of intermittent micro-plastic flow.

In the following section, the model is formally introduced and the DD technique described, and in sec.~\ref{sec_loading} various loading modes to produce a stress-strain curve are presented. Sec.~\ref{sec_results} presents the DD simulations for a wide range of model parameters to investigate their effect on deformation behaviour. The statistical analysis of the stress at which intermittent activity occurs, as well as the distribution of related strain-bursts, are presented in sec.~\ref{sec_statistics}. Finally, sec.~\ref{sec_discussion} discusses the context for the model, where it is argued that its applicable range is restricted to the micro-plastic region of the stress-strain curve --- a regime where significant structural evolution and work hardening are, to a large extent, absent~\cite{Young1961,Vellaikal1969}.

\section{Description of model} \label{sec_model}

\begin{figure}[t]
\centering
\includegraphics[clip,width=0.50\textwidth]{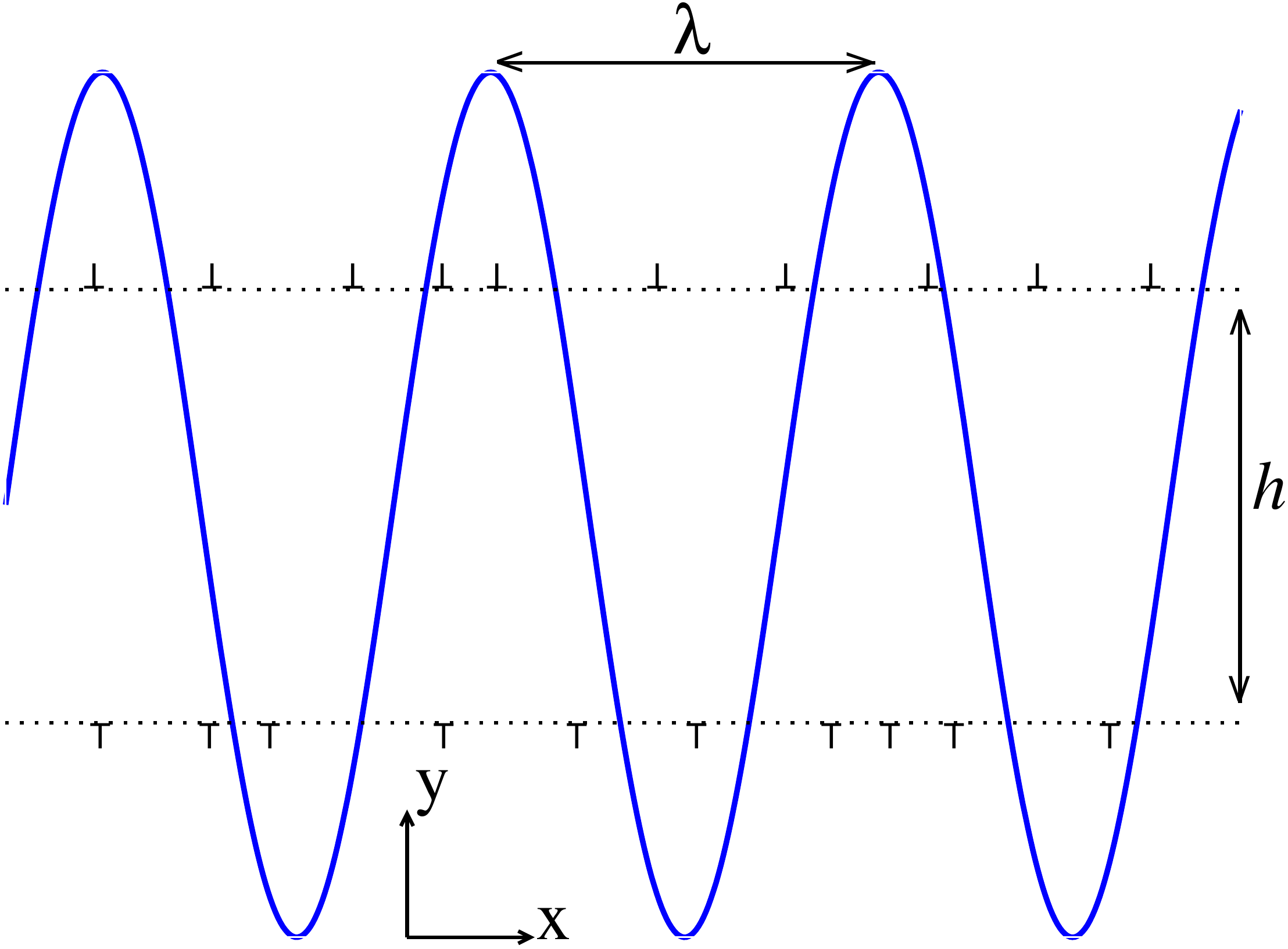}
\caption{\small Schematic of the two-dimensional dislocation dipolar mat system consisting of edge dislocations populating the two slip planes and the external sinusoidal stress field visualised in blue.}
\label{schematic_fig}
\end{figure}

The proposed model consists of two parallel slip planes, in the $x-z$ plane, separated by a distance $h$ along the $y$ direction. Each slip plane is populated by infinitely long straight edge dislocations where one slip plane contains $N_{+}$ dislocations with Burgers vector $\vec{b}=(b,0,0)$ and the other $N_{-}$ dislocations with Burgers vector $\vec{b}=(-b,0,0)$. Equal numbers of each type of dislocation are considered to ensure no net Burgers vector content: $N_{+}=N_{-}=N/2$. Such a structure is traditionally known as a simple dipolar mat~\cite{Hansen1986}. To this, a time independent internal sinusoidal shear-stress field is added of the form 
\begin{equation}
\tau_{\mathrm{Internal}}(x)=\tau_{0}\cos\left(\frac{2\pi x}{\lambda}\right), \label{EqInternal}
\end{equation}
which is parametrised by a shear-stress amplitude $\tau_{0}$ and a periodic length scale $\lambda$. Fig.~\ref{schematic_fig} displays a schematic of the model system in which the line direction of each straight dislocation is perpendicular to the plane of the figure. The dislocation density is defined by the total number of dislocations divided by the area of the system. The area of the system is defined by its length along $x$, defined as $d$, and its spatial extent along $y$, defined as $2h$. Detailed discussion on the origin of the internal sinusoidal stress field is deferred to sec.~\ref{sec_discussion}, however, its existence should be viewed as a simplified representation of a pre-existing (and unchanging) immobile dislocation population. The explicit dislocations, schematically shown in fig.~\ref{schematic_fig}, are therefore to be viewed as the mobile dislocation population of the system, and are quantified by their density $\rho_{\mathrm{m}}$.

In addition to the force arising from the internal shear-stress field, each dislocation will experience a force of elastic origin from all other dislocations within the system. Since it is assumed that the dislocations are unable to leave their slip plane, the only relevant force per unit dislocation length component will be along the $x$ direction and, via isotropic elasticity, may be calculated from the well known expression between two edge dislocations (labelled $i$ and $j$) on parallel slip planes~\cite{HullBaconBook}
\begin{equation}
f_{x,ij}=\frac{Gb_{x,i} b_{x,j}}{2\pi(1-\nu)}\frac{\Delta x(\Delta x^{2}-\Delta y^{2})}{(\Delta x^{2}+\Delta y^{2})^{2}}. \label{EqDDInt}
\end{equation}
Here $G$ is the isotropic shear modulus and $\nu$ Possion's ratio of the isotropic elastic medium, $b_{x,i}$ ($b_{x,j}$) is the Burgers vector in the x-direction of the $i$th ($j$th) dislocation, and $(\Delta x,\Delta y)$ is the two dimensional vector defining the dislocations' spatial separation. Presently a model isotropic Cu system is implemented, in which the shear modulus is taken as $G=42$ GPa, the Possion's ratio as $\mu=0.43$, and the Burgers vector magnitude as $b=2.55$ \AA.

For the present work, periodic boundary conditions along $d$, the dipolar mat direction, and open boundary conditions along $h$, are assumed. Due to the long range nature of eqn.~\ref{EqDDInt}, the correct treatment of periodicity invovles the summation of all dislocation image contributions to the force per unit dislocation length on a given dislocation. For the considered one-dimensional periodicity, an exact solution to such a summation is tractable, and is given by
\begin{eqnarray}
f_{x,ij}&=&-\frac{Gb_{x,i} b_{x,j}}{2(1-\nu)}\sin\left(\frac{2\pi\Delta x}{d}\right)\times \nonumber \\
& &\frac{\left[d\left(\cos\left(\frac{2\pi\Delta x}{d}\right)-\cosh\left(\frac{2\pi\Delta y}{d}\right)\right)+2\pi\Delta y\sin\left(\frac{2\pi\Delta y}{d}\right)\right]}{d^{2}\left(\cos\left(\frac{2\pi\Delta x}{d}\right)-\cosh\left(\frac{2\pi\Delta y}{d}\right)\right)^{2}}. \label{EqDDIntImage}
\end{eqnarray}
A simple derivation of this equation is detailed in appendix~\ref{appendix_a}.

The temporal evolution of a particular dislocation configuration is characterised by the choice of an empirical mobility law. Due to the actual discreteness of the lattice at the atomic scale, a dislocation segment must overcome an energy barrier associated with the local shearing of atoms in order to move an atomic distance. This so-called Peierls energy barrier and the associated Peierls stress~\cite{HullBaconBookRef}, the stress at which the dislocation can begin to move (defined at a given temperature), results in the dislocation moving quasi-statically from atomic lattice site to atomic lattice site. At the meso-scopic scale this results in over-damped motion where the dislocation's velocity is proportional to the force acting on the dislocation --- the present mobility law. The material specific proportionality constant is referred to as a damping parameter and is dependent on dislocation type, geometry, and on temperature. 

The equation of motion along the $x$ direction for the $i$th dislocation is then given by
\begin{equation}
\frac{\delta x_{i}}{\delta t}=\frac{F_{x,i}}{B}, \label{EqOfMotion}
\end{equation}
where $B$ is the damping coefficient, which for Cu is $5\times10^{-5}$ Pa s~\cite{Fournet1996,Devincre1993b}. In eqn.~\ref{EqOfMotion}, $F_{x,i}$, is the total force per unit dislocation length acting on the dislocation,
\begin{equation}
F_{x,i}=\left[\tau_{\mathrm{Internal}}(x_{i})-\tau_{\mathrm{External}}\right]b_{x,i}+\sum_{j\ne i}f_{x,ij}.\label{EqTotalForce}
\end{equation}
Here $\tau_{\mathrm{External}}$ is an externally applied homogeneous shear-stress field and $\tau_{\mathrm{Internal}}$ is the static sinusoidal shear-stress field defined in eqn.~\ref{EqInternal}. 

The numerical solution of eqn.~\ref{EqOfMotion} constitutes the Dislocation Dynamics (DD) algorithm presently used in which an appropriate finite time-step, $\delta t$, is used to integrate the equations of motion. The corresponding shear-strain response $\delta\varepsilon$ within this $\delta t$ is calculated via
\begin{equation}
\delta\varepsilon=\frac{1}{2dh}\sum_{i}^{N}b_{x,i}\delta x_{i}. \label{shear_strain_eqn}
\end{equation}
It is again emphasised that over the time-scale of $\delta t$ all atomic scale aspects are averaged over and inertial effects are ignored. Since the edge dislocations are infinitely long and straight such dynamics falls into the class of two dimensional DD modelling.

The dislocation density is given by $\rho_{\mathrm{m}}=\frac{N}{2dh}$. For the present work, the periodic length $d$ is chosen to define the distance $h$ between the two populated slip systems via $2d/N$. This sets the mean distance between dislocations along the $x-$ and $y-$direction to be the same, and defines the dislocation density as $\left(N/2d\right)^{2}$. Thus, $h=1/\sqrt{\rho_{\mathrm{m}}}$, and choosing a value of the dislocation density will fix the scale of interaction between the two parallel slip planes of the dipolar mat geometry --- see appendix A. The motivation for such a restriction is to give the dislocation density a greater bulk-like relevance, where in the bulk limit it represents an isotropic density of dislocations. Other definitions of $h$ are, of course, possible.

\section{Loading and the calculation of stress-strain curves} \label{sec_loading}

To simulate a loading experiment and thus a stress-strain curve, a sample must be produced and a loading mode chosen. Presently, sample preparation involves the chosen number of dislocations being initially placed at random positions within the dipolar mat geometry, and the structure relaxed via eqn.~\ref{EqOfMotion} to minimise the force on each dislocation to within a chosen tolerance. This is performed at $\tau_{\mathrm{External}}=0$. In the present work, no attempt is therefore made to determine a low (or lowest) energy dislocation structure, an approach that is compatible with the fact that the explicit dislocations within the model are only those of the mobile variety. This aspect is found to be crucial to the properties of the model and will be discussed in more detail in sec.~\ref{sec_discussion}.

There exist a number of ways in which a deformation simulation can be done. The first such loading mode is referred to as ``stress-relaxed'' and involves incrementing the external shear-stress by a value $\delta\tau$ ($\tau_{\mathrm{External}}\rightarrow\tau_{\mathrm{External}}+\delta\tau$), and relaxing the structure until the the sum of the dislocation force magnitudes, or equivalently, the dislocation velocity magnitudes, varies by less than the fraction $10^{-8}$.

\begin{figure}[h]
\centering
\includegraphics[clip=true,width=0.95\textwidth]{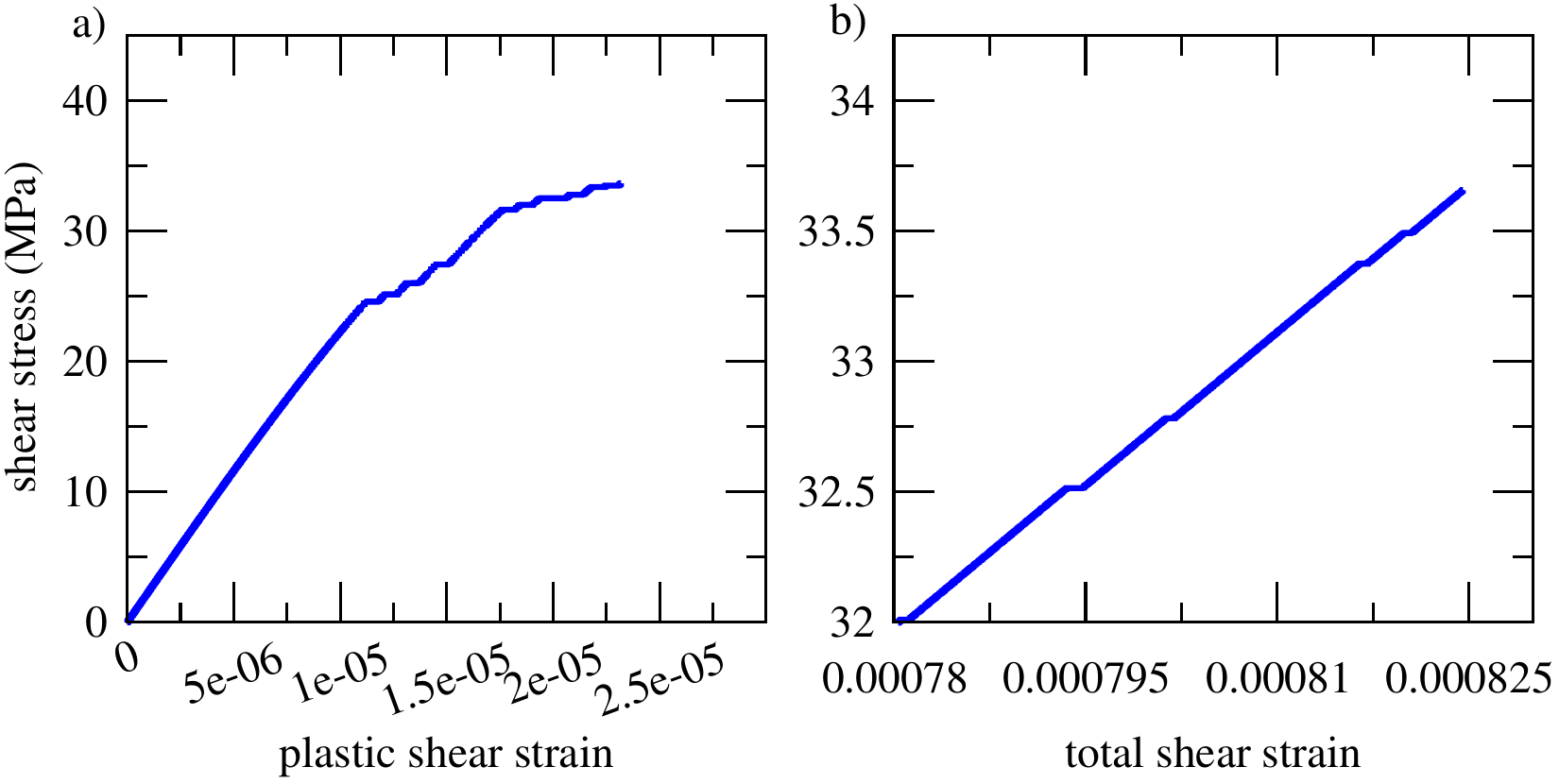}
\caption{\small a) A typical stress versus plastic strain curve produced using the ``stress relaxed'' loading mode. Stress plateaus indicating strain burst activity is clearly evident. b) Displays, for the same simulation, the corresponding stress versus total (elastic plus plastic) strain for a small range of shear-stresses.} \label{fig_stress_relaxed_ss_example}
\end{figure}

During the relaxation, the associated plastic shear-strain increment may be calculated as the sum of eqn.~\ref{shear_strain_eqn} over all $\delta t$ time steps of the relaxation. Once convergence is obtained this shear-strain increment is added to the total plastic shear-strain and the cycle is repeated until the desired stress-strain curve is obtained. Fig.~\ref{fig_stress_relaxed_ss_example}a displays a typical shear-stress versus plastic shear-strain curve during the initial stages of loading. Discrete dislocation activity, via strain bursts, is clearly evident and is separated by continuous an-elastic regions. By adding the elastic shear-strain, $\tau/G$, to the plastic shear-strain, the total shear-strain is obtained. Fig.~\ref{fig_stress_relaxed_ss_example}b displays the corresponding shear-stress versus total shear-strain curve for a small range of stresses and an approximately linear stress-strain curve with intermittent stress plateaus is seen. When displaying a similar curve for the full range of stresses, as seen in fig.~\ref{fig_stress_relaxed_ss_example}a, only a straight line is resolvable indicating that such strain bursts are well within the micro-plastic regime of deformation. At larger stresses a plastic flow regime is entered which will be investigated in more detail in subsequent sections.

Experimentally, two distinct deformation modes can be used: displacement controlled and load controlled. Here we consider an inherently force controlled testing device. In such a case, displacement controlled testing is done by adjusting the applied load via a feed-back loop such that the displacement rate is held at a fixed value throughout the loading, whereas for load controlled experiments, the applied load simply increases at a chosen rate. For the present model, these deformation modes correspond to a constant shear-strain rate and constant shear-stress rate loading condition. To obtain a stress-strain curve with a constant shear-stress rate, a numerical value for the applied stress rate, $\dot{\tau}_{\mathrm{External}}$, is chosen. This then defines a stress increment $\delta\tau=\dot{\tau}_{\mathrm{External}}\delta t$, where $\delta t$ is the time-step used to evolve the dislocation network according to eqn.~\ref{EqOfMotion}. Thus at every simulation iteration, the stress is increased by $\delta\tau$ and the configuration evolves in time by an amount $\delta t$. To implement a constant shear-strain rate loading mode, $\dot{\varepsilon}$, a numerical value is chosen and the appropriate $\delta\tau$ stress increment, to achieve such a strain rate, is performed every simulation step. The actual value of $\delta\tau$ is determined by assuming that the total strain rate decomposes additively into an elastic and plastic component:
\begin{equation}
\dot{\varepsilon}=\dot{\varepsilon}_{\mathrm{elastic}}+\dot{\varepsilon}_{\mathrm{plastic}}=\frac{\dot{\tau}_{\mathrm{External}}}{G}+\dot{\varepsilon}_{\mathrm{plastic}}, \label{EqConstStrainRate}
\end{equation}
giving
\begin{equation}
\dot{\tau}_{\mathrm{External}}=G\left(\dot{\varepsilon}-\dot{\varepsilon}_{\mathrm{plastic}}\right)
\end{equation}
or
\begin{equation}
\delta\tau=G\left(\dot{\varepsilon}-\dot{\varepsilon}_{\mathrm{plastic}}\right)\delta t, \label{ConstSR}
\end{equation}
where $\dot{\varepsilon}_{\mathrm{plastic}}\delta t$ is the strain increment per simulation iteration, calculated via eqn.~\ref{shear_strain_eqn}. Thus the stress increment $\delta\tau$ is determined by the correction needed to achieve the required constant strain rate for the next simulation iteration. In the above, due to the simplified geometry of the model (fig.~\ref{schematic_fig}), a pure shear modulus, $G$, rather than a Young's modulus is used.

\begin{figure}[h]
\centering
\includegraphics[clip=true,width=0.95\textwidth]{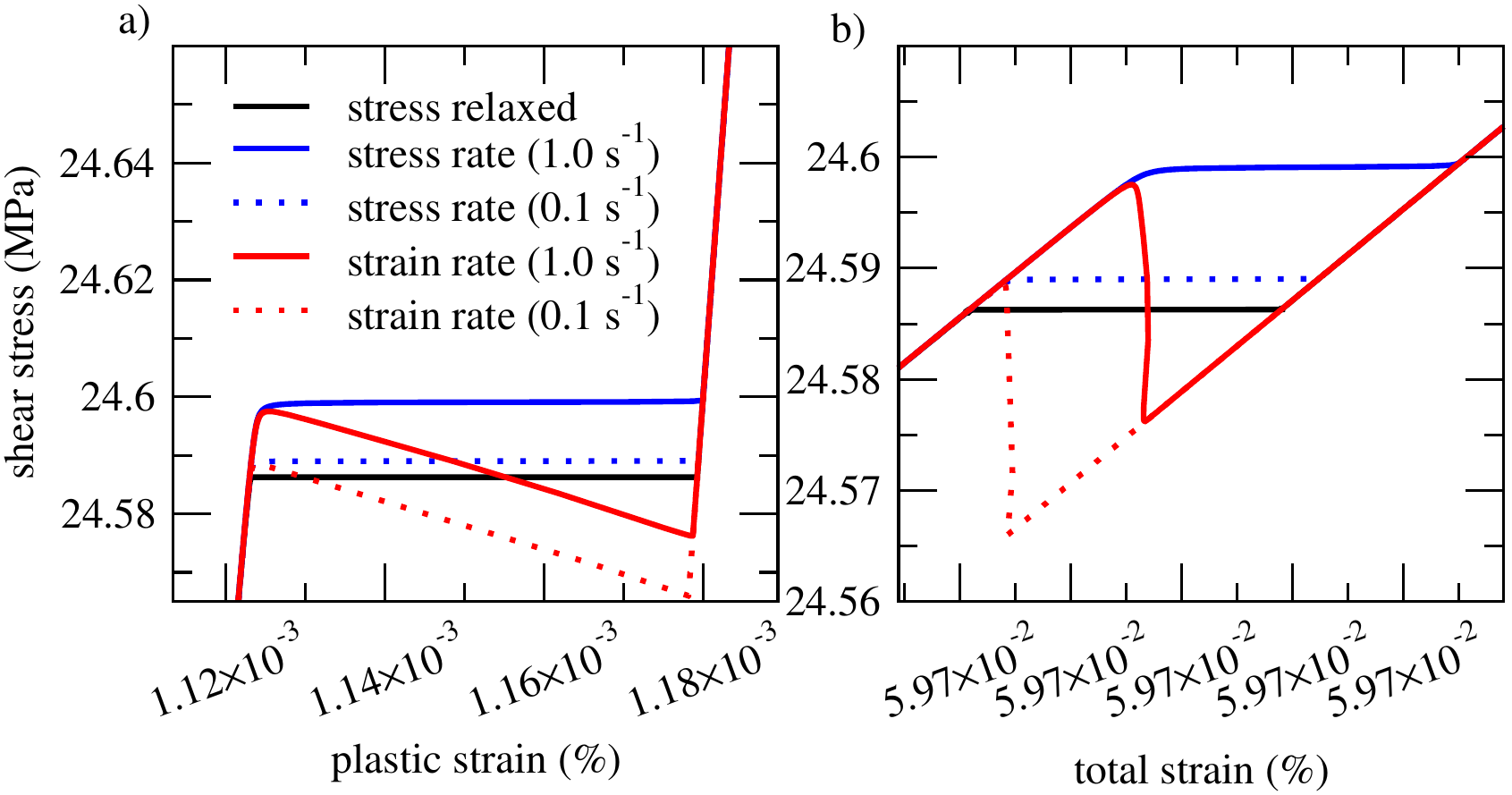}
\caption{\small Close up of a dislocation burst for the three considered loading modes and at different strain rates. a) Stress versus plastic strain and b) stress versus total (elastic plus plastic) strain.}
\label{fig_burst_modes}
\end{figure}

\begin{figure}[h]
\centering
\includegraphics[clip=true,width=0.95\textwidth]{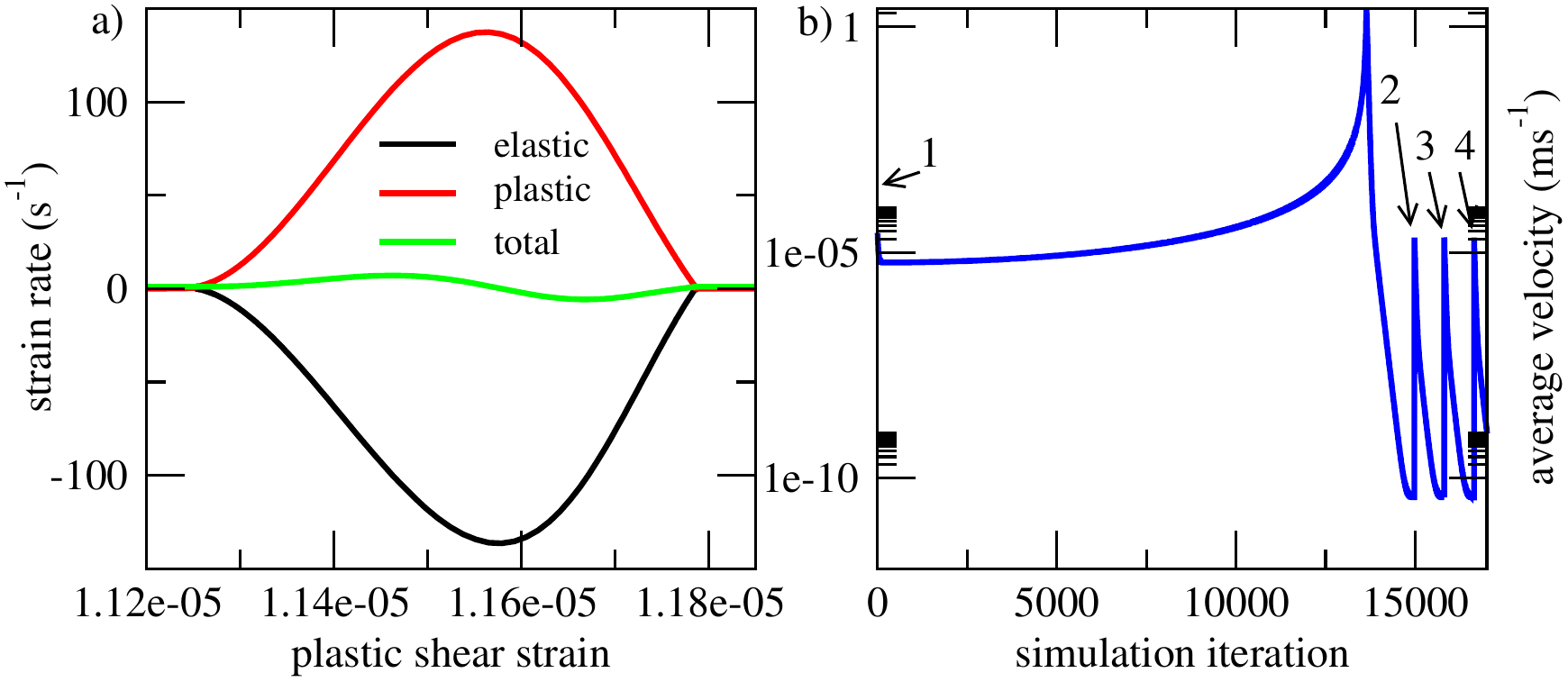}
\caption{\small a) Elastic and plastic strain rates as a function of strain for the dislocation burst displayed in fig.~\ref{fig_burst_modes}. b) Average dislocation velocity during the same dislocation burst in ``stress relaxed'' deformation mode.}
\label{fig_burst_modes_strain_rate}
\end{figure}

To investigate how such loading modes affect the discrete dislocation activity seen in fig.~\ref{fig_stress_relaxed_ss_example}, the appropriate stress-rate and strain-rate must be chosen. This is done by first choosing $\dot{\varepsilon}$, from which $\dot{\tau}_{\mathrm{External}}$ is obtained via $\dot{\varepsilon}/G$ to ensure that in the elastic/an-elastic regime both deformation modes have the same total strain rate. Fig.~\ref{fig_burst_modes} displays a single strain burst in all three considered loading modes. For the constant strain rate and stress rate modes two strain rates are considered: 0.1 s$^{-1}$ and 1.0 s$^{-1}$. For the ``stress-relaxed'' loading mode a sharp plateau is evident in fig.~\ref{fig_burst_modes}a with an identically zero gradient. In this region, the constant stress rate mode also exhibits a plateau but with a (non-zero) positive gradient since, during the evolution of the strain burst, the stress is rising at the chosen rate. Also, the stress at which the strain burst initiates is somewhat higher (and increasing with increasing strain rate) than that in the stress relaxed mode indicating a strain-rate effect. In this regard, the ``stress-relaxed'' deformation mode can be considered as the zero stress-rate limit of the constant stress rate loading mode in which the dislocation configuration always has time to relax before the next stress increment. For the constant strain-rate loading mode, the onset of the strain burst occurs at similar stresses to that of the constant stress rate, however as the strain-burst evolves, the stress decreases to maintain the chosen strain-rate. Fig.~\ref{fig_burst_modes}b displays the same strain burst with the stress now as a function of the total strain. The greatest effect is seen in the constant strain-rate mode where due to the drop in stress during the strain-burst, there is a rapid drop in elastic strain. Fig.~\ref{fig_burst_modes_strain_rate}a displays the plastic strain rate as a function of plastic strain for the burst shown in fig.~\ref{fig_burst_modes} (for a constant strain rate of 1.0 s$^{-1}$). The  elastic strain rate response as dictated by eqn.~\ref{ConstSR} is also shown and correspondingly reduces to compensate for the rise in the plastic strain rate. Data for a constant strain rate of 0.1 s$^{-1}$ differs little from the 1.0 s$^{-1}$ data, a result (along with the high strain-rates of fig.~\ref{fig_burst_modes_strain_rate}a) due to the driven zero temperature nature of the simulation.

It is noted that for the constant strain rate loading mode, the loading system responds instantaneously (to within $\delta t$) to any discrete plastic event (hence leading to the backwards curvature seen in fig.~\ref{fig_burst_modes}b). To model an instrumentally realistic device, a more complex differential equation than that of eqn.~\ref{ConstSR} would need to be developed, which takes into account the finite delay time and resolution of the loading apparatus.

Fig.~\ref{fig_burst_modes_strain_rate}b now shows the average dislocation velocity during the ``stress relaxed'' deformation mode for four consecutive applied shear stresses in which the first constant stress relaxation undergoes the irreversible plastic strain rate seen in fig.~\ref{fig_burst_modes}. For this stress value, an initial relaxation of the dislocation velocities is seen, followed by a rapid acceleration of the dislocations corresponding to the emerging instability and the eventual irreversible plastic event of fig.~\ref{fig_burst_modes}. This continues until a maximum velocity is reached at which the dislocations then begin to decelerate until the convergence criterion is met. Three further constant stress increments result in an immediate relaxation of the dislocation velocities corresponding to a reversible (anelastic) relaxation of the dislocation configuration.

\section{Results} \label{sec_results}

In this section the ``stress-relaxed'' loading mode is used to investigate the influence of the model parameters on the stress-strain curve.

\subsection{Stress-strain behaviour as a function of micro-structural parameters: $\tau_{0}$, $\lambda$, and the mobile dislocation density, $\rho_{\mathrm{m}}$.}

\begin{figure}[h]
\centering
\includegraphics[clip=true,width=0.95\textwidth]{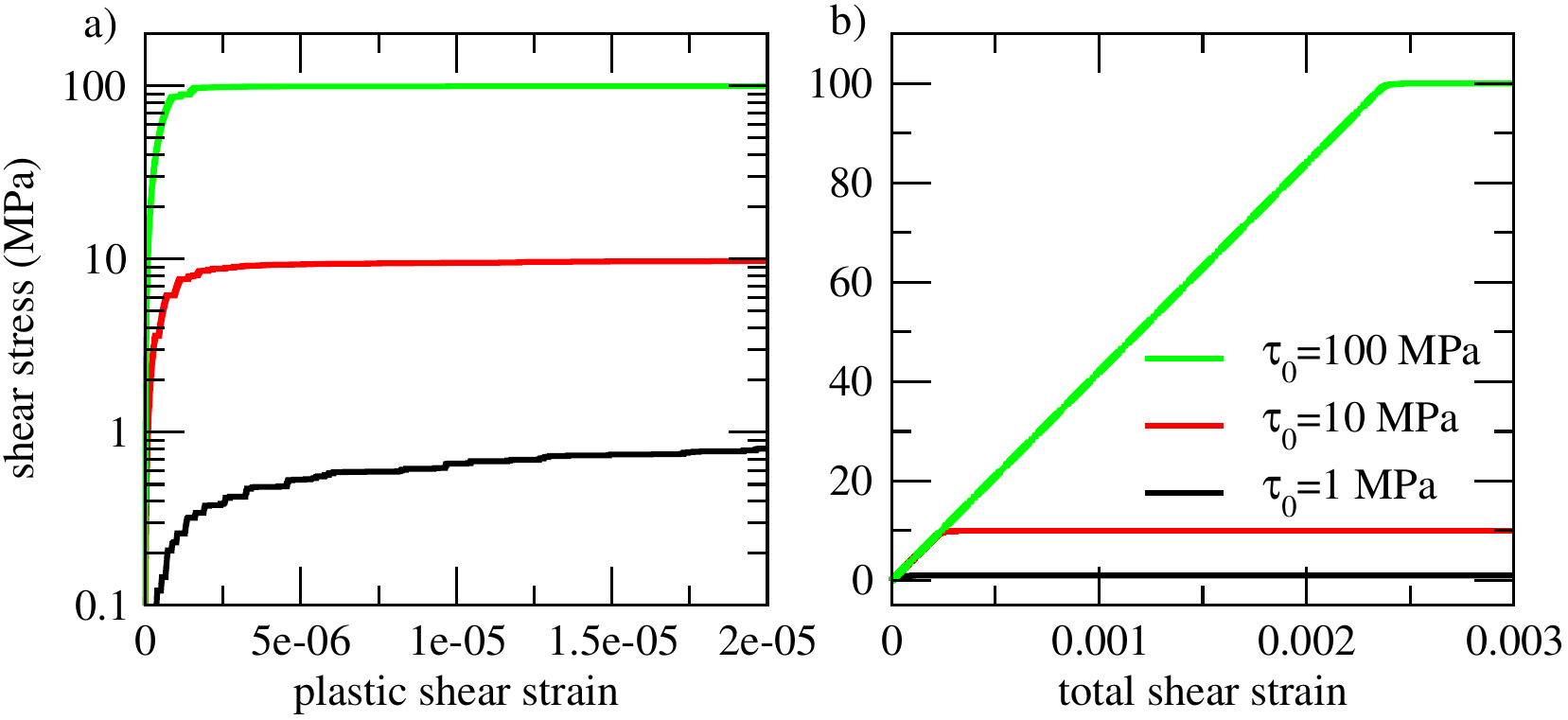}
\caption{\small a) Stress versus plastic strain for three different $\tau_{0}$ values, for a mobile dislocation density $1\times10^{10}\mbox{m}^{-2}$, and b) the corresponding stress versus total strain.} \label{different_tau_fig}
\end{figure}

The influence of the sinusoidal shear-stress field amplitude is first investigated. Fig.~\ref{different_tau_fig}a displays the resulting shear-stress versus plastic shear-strain curves for three different choices of $\tau_{0}$; namely 100MPa, 10MPa and 1MPa for a system with $d=200$ $\mu$m and $\lambda=2$ $\mu$m, containing $N=N_{+}+N_{-}=20+20$ dislocations. This gives a mobile dislocation density of $\rho_{\mathrm{m}}=1\times10^{10}\mbox{m}^{-2}$. The figure demonstrates that the choice of $\tau_{0}$ strongly controls the stress at which macroscopic plastic flow occurs. In fig.~\ref{different_tau_fig}a, the vertical axis is plotted as a logarithmic scale to reveal the fine structure of the micro-plastic region. In all cases discrete strain bursts are evident. Fig.~\ref{different_tau_fig}b displays the corresponding shear-stress versus total shear-strain curves. The small plastic shear-strain values evident in figure~\ref{different_tau_fig}a and the sharp yield transition in \ref{different_tau_fig}b emphasise that the shear-stress versus plastic shear-strain data prior to flow is clearly in the micro-plastic regime of the stress-strain curve.

\begin{figure}[h]
\centering
\includegraphics[clip=true,width=0.95\textwidth]{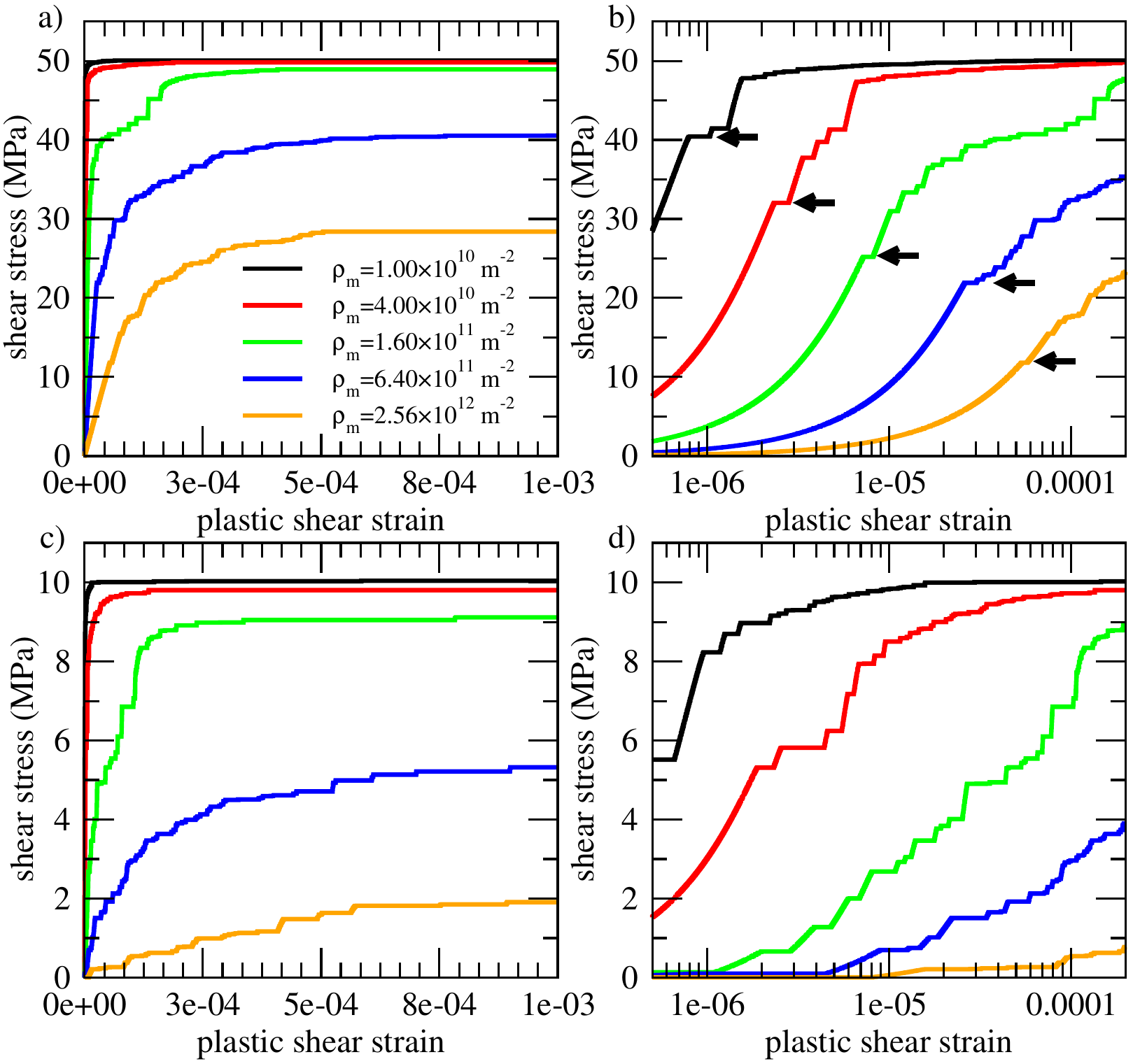}
\caption{\small  Stress versus plastic strain for different mobile dislocation densities for (upper panels) $\tau_{0}=50$ MPa and (lower panels) $\tau_{0}=10$ MPa. The right panels display the same data with strain plotted using a logarithmic scale to emphasise the initial micro-plastic strain region where the arrows (in b) indicate the first occurrence of a strain response.} \label{different_rho_fig}
\end{figure}

In figs.~\ref{different_rho_fig}a-b, $\rho_{\mathrm{m}}$ is now varied from $1\times10^{10}\mbox{m}^{-2}$ to $2.56\times10^{12}\mbox{m}^{-2}$ by changing the number of dislocations for a fixed $d=100$ $\mu$m and $\lambda=2$ $\mu$m. Here $\tau_{0}$ is 50 MPa. Figs.~\ref{different_rho_fig}a-b demonstrate that with increasing mobile dislocation density the flow stress decreases from $\tau_{0}$. Thus at low enough dislocation densities (as in fig.~\ref{different_tau_fig}), $\tau_{0}$ defines the flow stress of the system, a natural result since a well isolated dislocation will be primarily affected by the sinusoidal shear-stress field. Figs.~\ref{different_rho_fig}a-b also demonstrate that the micro-plastic regime broadens with increasing dislocation density and that the first discrete dislocation event occurs at a decreasing applied homogeneous shear-stress (see arrowed regions in fig.~\ref{different_rho_fig}b). Figs.~\ref{different_rho_fig}c-d show similar data for  $\tau_{0}=10$ MPa, for the same values of dislocation density.  Although a similar trend is seen, the deviation away from the maximum flow stress of $\tau_{0}=10$ is a proportionally stronger function of the increasing dislocation density.

\begin{figure}[h]
\centering
\includegraphics[clip=true,width=0.95\textwidth]{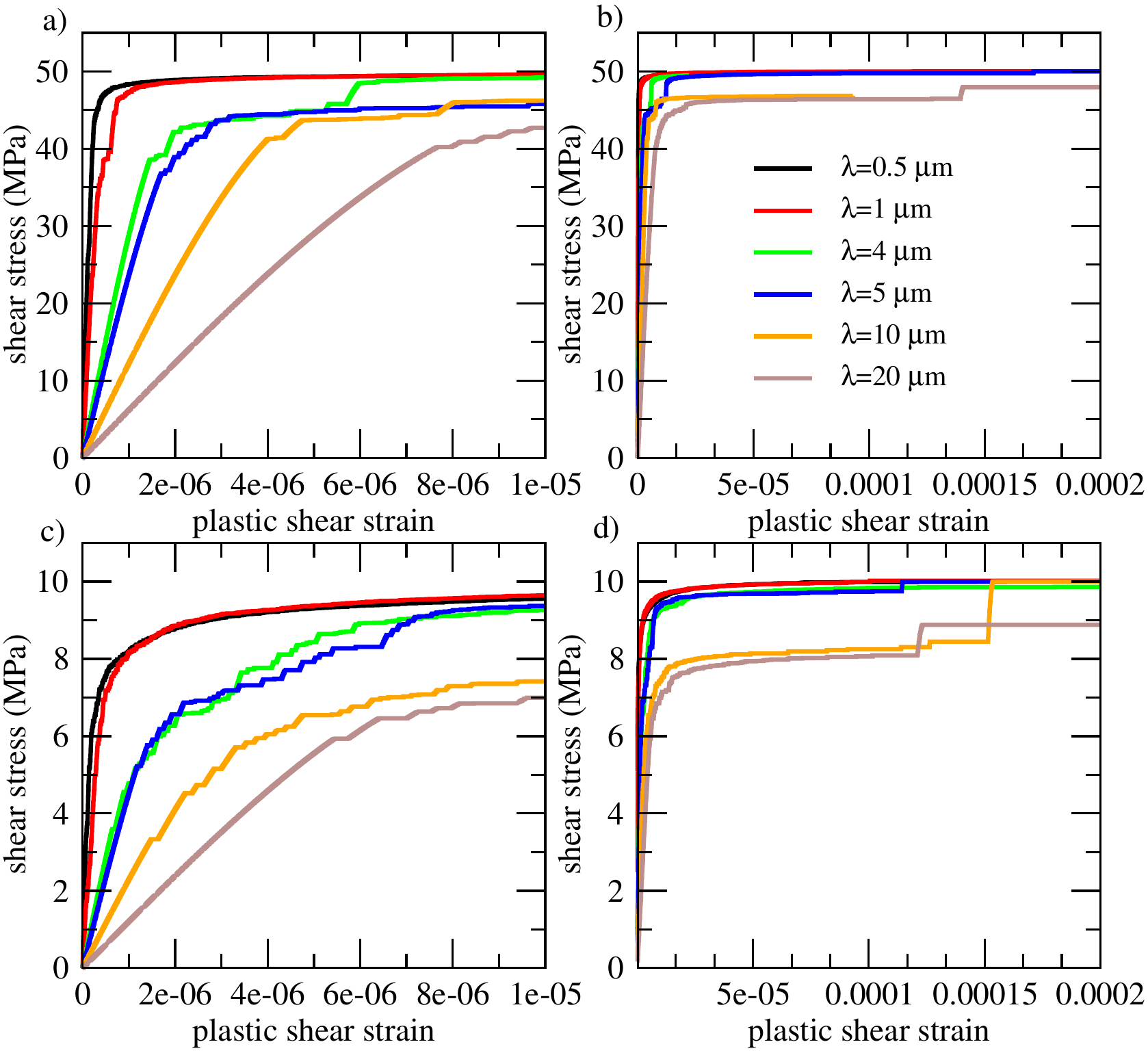}
\caption{\small Stress versus plastic strain for a range of $\lambda$ choices, where for each simulation $\tau_{0}=50$ MPa (upper panels) and $\tau_{0}=10$ MPa (lower panels), $d=800\mu$m and a mobile dislocation density $1\times10^{10}\mbox{m}^{-2}$.  The left panels show the initial strain regime and the right panels show a larger strain range.}
\label{different_lambda_fig}
\end{figure}

The remaining micro-structural variable that can be varied is the characteristic length scale of the sinusoidal stress field, $\lambda$. Fig.~\ref{different_lambda_fig} displays stress versus plastic strain curves as $\lambda$ increases from $\lambda=0.5\mu$m to $\lambda=20\mu$m, with $\tau_{0}=10$ MPa and 50 MPa, $d=800\mu$m and $\rho_{\mathrm{m}}=1\times10^{10}\mbox{m}^{-2}$. Here a larger $d$ was used to obtain a better statistical sample of the dislocation environment at the length scale of $\lambda$. The chosen dislocation density is similar to that of fig.~\ref{different_tau_fig} in which $\tau_{0}$ largely determined the flow stress. Inspection of fig.~\ref{different_lambda_fig} reveals that with increasing $\lambda$ the micro-plastic regime is broadened due to increasing strain burst magnitudes with increasing $\lambda$. For small strain (figs.~\ref{different_lambda_fig}a and \ref{different_lambda_fig}c), it also appears that the flow stress regime occurs at a reduced stress (for increasing $\lambda$), however at larger strains (figs.~\ref{different_lambda_fig}b and \ref{different_lambda_fig}d) flow stresses approximately equal to $\tau_{0}$ are eventually reached. It is also noted that with increasing $\lambda$ the strain burst magnitude increases. With decreasing $\tau_{0}$, the deviation of the curves away from $\tau_{0}$ is a stronger function of increasing $\lambda$ resulting in a somewhat broader micro-plastic region.

The central properties of the present model are reflected in figures~\ref{different_tau_fig}, \ref{different_rho_fig} and \ref{different_lambda_fig}. If the dislocation density is low enough or $\tau_{0}$ sufficiently high, then $\tau_{0}$ will primarily determine the shear-stress at which the extended plastic flow regime begins. In this limit, the internal shear-stress that each dislocation experiences is due to the sinusoidal stress field and when $\tau_{\mathrm{applied}}$ approaches $\tau_{0}$, the plastic flow regime is immediately encountered with a negligible micro-plastic region. By increasing the mobile dislocation density or decreasing $\tau_{0}$ the micro-plastic regime broadens and also the stress at which extended plastic flow occurs decreases due to the increasing influence of the elastic interaction between the dislocations. This latter feature is also reflected in the decreasing stress at which the first strain burst is seen. By increasing $\lambda$, the strain burst magnitude increases which broadens the micro-plastic regime.

\subsection{Stain burst behaviour as a function of periodic length scale} \label{sec_d_dependence}

\begin{figure}[h]
\centering
\includegraphics[clip=true,width=0.95\textwidth]{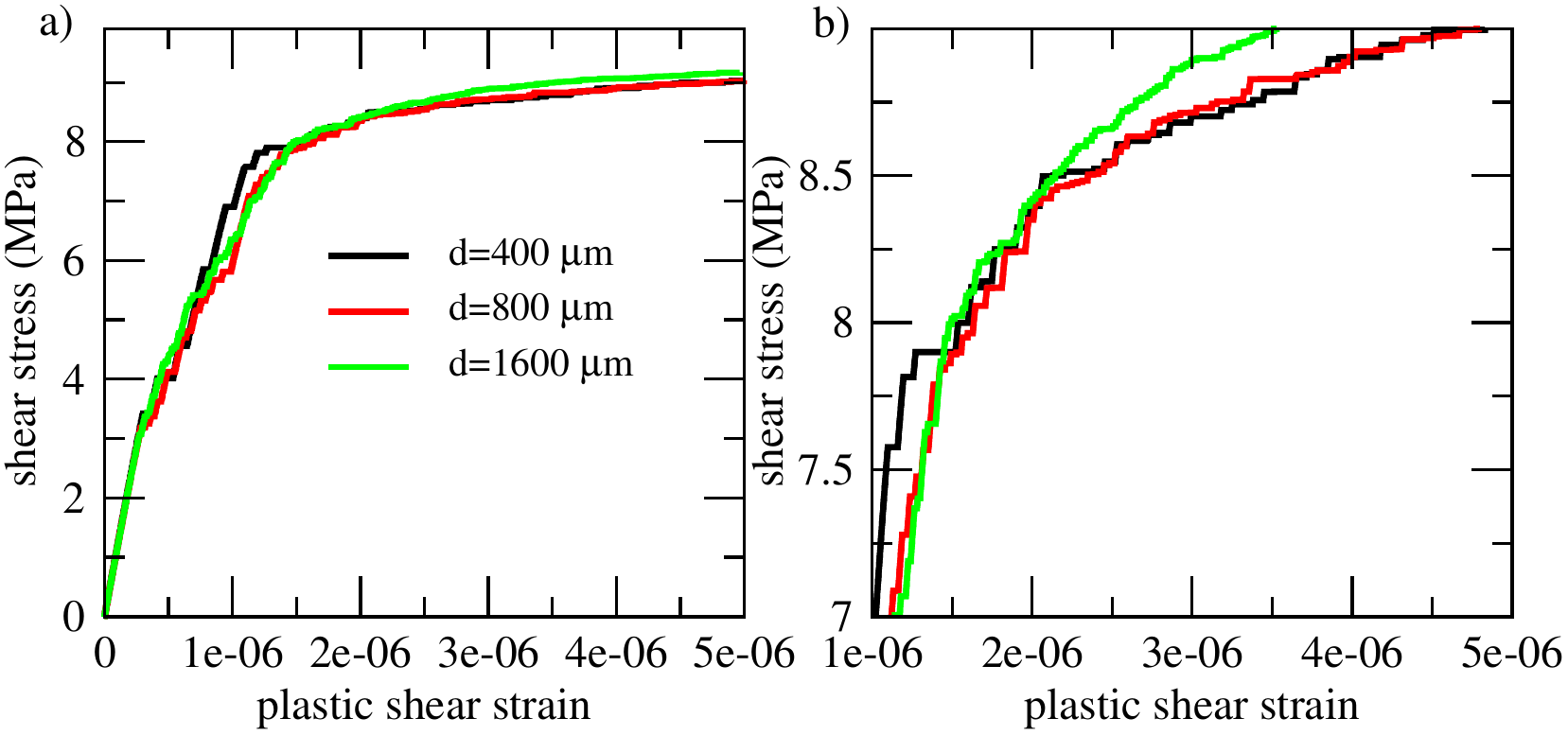}
\caption{\small a) Early stress versus plastic strain for different periodicity lengths $d$ with $\tau_{0}=10$ MPa and a mobile dislocation density $1\times10^{10}\mbox{m}^{-2}$. b) A close up of the stress-stain curves indicating that with increasing periodicity the minimum discrete strain bursts reduce in strain magnitude.}
\label{different_d_fig}
\end{figure}

The stress versus plastic strain behaviour for three different values of $d$, using $\tau_{0}=10$ MPa and $\rho_{\mathrm{m}}$ equal to $1\times10^{10}\mbox{m}^{-2}$ is presently investigated. For all three simulations $\lambda=2$ $\mu$m. Fig.~\ref{different_d_fig} displays the resulting stress-strain curves demonstrating that there is no strong overall $d$ dependence. Indeed, the stress at which the first strain burst occurs differs little for the three samples: $3.4$ MPa for $d=400\mu\mbox{m}$, $3.2$ MPa for $d=800\mu\mbox{m}$ and $3.0$ MPa for $d=1600\mu\mbox{m}$. It is seen, however, that with increasing $d$ the scale of the discrete plastic strain bursts becomes finer, leading to small-scale differences in the curves. Inspection of the minimum magnitude of the plastic strain bursts observed in this figure, and in figures~\ref{different_tau_fig} and \ref{different_rho_fig}, reveals it to be approximately $\left|b\right|\lambda/2dh$, indicating that the strain bursts correspond to dislocation motion over a distance equal to at least $\lambda$. It is worth noting that the existence of such {\em minimum} discrete strain bursts is an artifact of the periodicity used to approximate a system of infinite extent. Indeed, in the limit $d\rightarrow\infty$ the minimum strain bursts would approach zero. 

\section{Statistical properties of intermittent flow} \label{sec_statistics}

\subsection{First burst behaviour}

\begin{figure}[h]
\centering
\includegraphics[clip=true,width=0.95\textwidth]{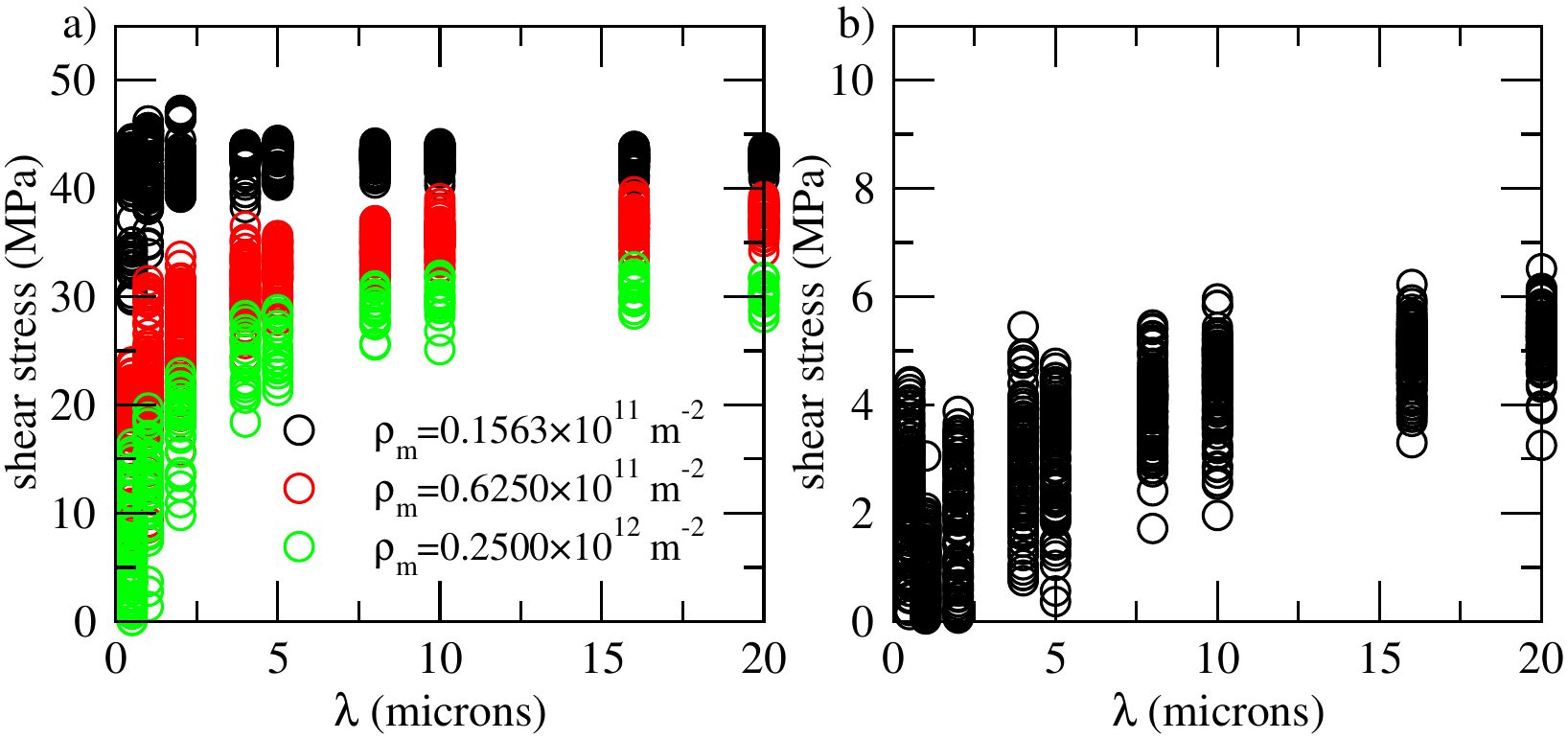}
\caption{\small a) Plot of stress of first strain burst as a function of $\lambda$ for three dislocation densities for $\tau_{0}=50$ MPa and b) $\tau_{0}=10$ MPa at a dislocation density of $0.1563\times10^{10}\mbox{m}^{-2}$.}
\label{first_burst_fig}
\end{figure}

To better understand the relationship between the mobile dislocation density and $\lambda$, a series of simulations is performed to investigate the statistics of the shear-stress at which the first strain burst occurs. For computational efficiency these simulations were done using the constant stress-rate loading mode with $d=800$ $\mu$m and $\tau_{0}=50$ MPa. Fig.~\ref{first_burst_fig}a displays such shear-stress values as a function of $\lambda$ for three different mobile dislocation densities. For each $\lambda$ and dislocation density value, the results of many different simulations are shown to indicate the degree of statistical variation. Inspection reveals that by increasing the dislocation density, the first-burst stress value becomes increasingly dependent on the value of $\lambda$, decreasing with decreasing $\lambda$. For all three considered dislocation densities, the degree of scatter increases with decreasing $\lambda$.

The state of the dislocation configuration prior to loading provides an understanding of the results of fig.~\ref{first_burst_fig}a: $N$ dislocations are introduced into the dipolar mat by randomly placing them along the two possible slip planes and the system is then relaxed to a local equilibrium configuration. For large enough $\lambda$ this relaxation process will result in similar numbers of dislocations for each $\lambda$ unit, producing a dislocation configuration that is increasingly ordered. However, as $\lambda$ decreases, there will exist an increasingly varying amount of dislocations for each $\lambda$ unit, and the original disorder associated with the random initial positions of the dislocations is increasingly preserved. These two trends constitute the origin of the increased scatter and general decrease seen in fig.~\ref{first_burst_fig}a with respect to decreasing $\lambda$, since the first-burst stress value is directly related to the lowest critical shear-stress required for an irreversible configurational change to occur. That is, the first-burst stress probes the extremal values of the dislocation environment and a greater variation in dislocation environment will naturally lead to a greater variation in the first-burst stress, which also increases scatter and results in an on-average reduced stress magnitude. Dislocation interaction enhances this effect by generally increasing the degree of disorder in the low-lambda limit.

The above result is, in fact, dependent on the system size (the periodicity length, $d$) being finite. As the system size grows at a fixed value of $\lambda$ and dislocation density, the chances for an extremal dislocation configuration to occur increases, resulting in a increased scatter and reduction of the first-burst shear-stress. It is therefore expected that as the system size increases, the saturation in first burst stresses as a function of increasing $\lambda$ will be shifted to larger values of $\lambda$. Such a size effect must also be reflected in the choice of $\tau_{0}$ which, along with the dislocation interaction, determines the degree of configurational disorder in the initial state. For a given value of $\lambda$, a smaller value of $\tau_{0}$ should increase the disorder due to the increasing influence of the dislocation interaction on the initial configuration. Fig.~\ref{first_burst_fig}b displays a similar figure of first burst stresses as a function $\lambda$ using a value of $\tau_{0}=10$ MPa and, indeed, demonstrates a shift to the right of the saturation region.

Although these results are an artifact of the system size, they reveal that through proper choice of model parameters, the intermittent plasticity seen in the past sections is driven by the extreme value statistics associated with the initial dislocation configuration --- a behaviour that is to be equally expected in real crystals.

\subsection{Distribution of strain burst magnitudes} \label{sec_statistics_dist}

\begin{figure}[h]
\centering
\includegraphics[clip=true,width=0.95\textwidth]{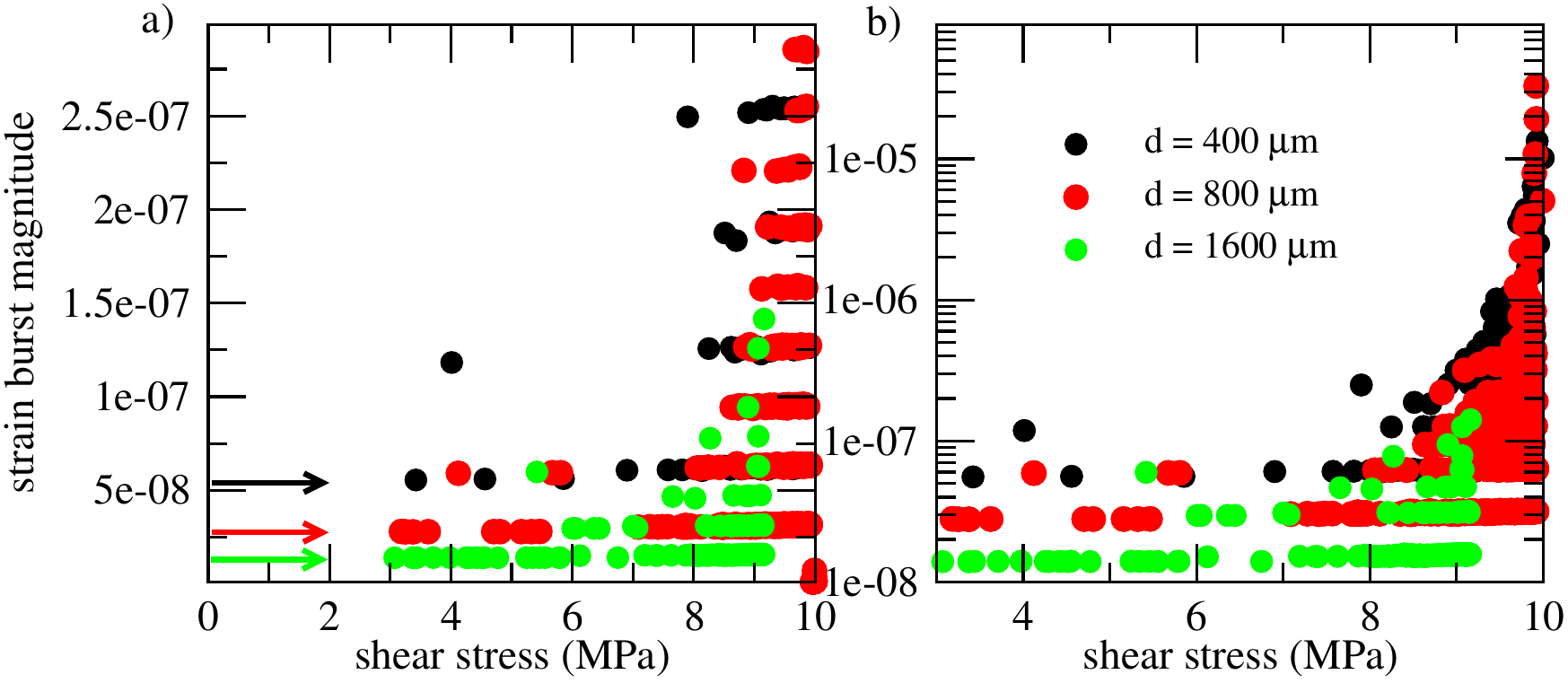}
\caption{\small a) Magnitude of strain bursts as a function of the applied shear-stress for three samples differing only in their size $d$. The arrows indicate the (minimum) fundamental strain bursts for each sample, indicating that larger strain bursts occur as multiples of these values. b) Vertical log plot for the entire applied stress range demonstrating that as the flow stress regime is approached individual strain bursts can become large.}
\label{strain_burst_fig}
\end{figure}

The statistical properties of the strain burst magnitudes are now investigated for a system with $\tau_{0}=10$ MPa, $\lambda=2$ $\mu$m, and $\rho_{\mathrm{m}}$ equal to $1\times10^{10}\mbox{m}^{-2}$. Three systems with $d=400$, $800$  and $1600$ $\mu$m are first considered. The stress-strain curves for the $d=400\mu\mbox{m}$ and $d=800\mu\mbox{m}$ systems span up to the flow stress regime at approximately $\tau_{0}=10$ MPa, whereas for the $d=1600\mu\mbox{m}$ data is only available up to approximately 9.4 MPa. Fig.~\ref{strain_burst_fig} plots the magnitude of the strain burst as a function of the shear-stress at which it occurred, for all three samples. In the first instance, the plastic strain magnitudes for each stress step (as calculated via eqn. 6) were analysed and it was found that irreversible plasticity via discrete strain bursts begins at approximately 3 MPa with magnitudes equal to $\left|b\right|\lambda/2dh$ (indicated by the corresponding coloured arrows). With increasing stress, larger strain bursts occur with magnitudes equal to multiples of the basic strain burst unit for each sample. Fig.~\ref{strain_burst_fig}b displays the entire range of strain bursts using a vertical log-scale, and it is seen that the spread in strain bursts increases rapidly as the shear-stress approaches the flow regime at approximately $\tau_{0}=10$ MPa.

\begin{figure}[h]
\centering
\includegraphics[clip=true,width=0.95\textwidth]{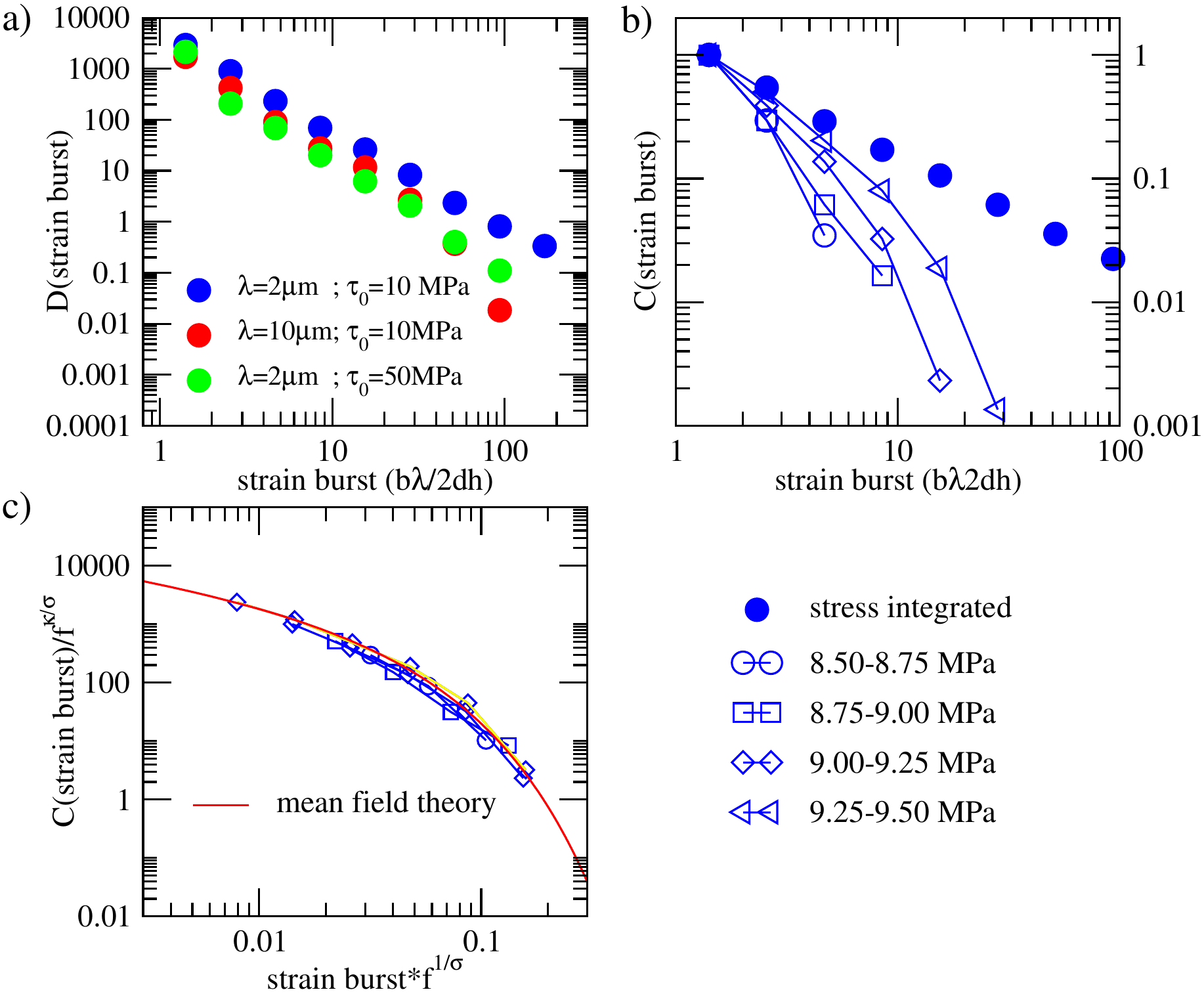}
\caption{\small a) Statistics of strain bursts derived from stress-strain simulations for three choices of parameters. Data is shown as a log-log plot of the histogram of strain bursts in units of the fundamental strain burst magnitude. b) Corresponding stress dependent and stress integrated complementary cumulative distribution for the case of $\lambda=2$ $\mu$m and $\tau_{0}=10$ MPa and c) via an appropriate scaling procedure where $f=(\tau_{0}-\tau)/\tau_{0}$ all stress dependent complementary cumulative distributions collapse on the universal curve given by mean field theory --- see ref.~\cite{Dahmen2009}.}
\label{strain_burst_dist1_fig}
\end{figure}

Plotting the strain-burst magnitude distribution derived from the data of fig.~\ref{strain_burst_fig}, reveals a power law behaviour which is similar for each of the three considered systems. However from these graphs (not shown),  a reliable exponent could not be obtained and to investigate this aspect further improved statistics is required.  Fig.~\ref{strain_burst_dist1_fig}a shows strain-burst magnitude distributions from a much larger number of events, plotted using a log-log scale, for three different choices of parameters: 1) $\lambda=2$ $\mu$m and $\tau_{0}=10$ MPa (same as that of fig.~\ref{strain_burst_fig}), 2) $\lambda=10$ $\mu$m and $\tau_{0}=10$ MPa, and 3) $\lambda=2$ $\mu$m and $\tau_{0}=50$ MPa. For all simulations $d=800$ $\mu$m and $\rho_{\mathrm{m}}=1\times10^{10}$ m$^{2}$. For 1), eleven deformation simulations were performed, and for 2) and 3) sixteen independent simulations were performed for each system. All distributions were obtained by logarithmic binning of identified irreversible plastic strain events. Inspection of this figure reveals good power law behaviour over a number of decades. This is particularly the case for the parameter set 1), in which more statistics could be obtained from each individual curve. Indeed, for this parameter-set, fig.~\ref{first_burst_fig} shows that the first-burst stresses are low and exhibit strong scatter indicating significant intermittent plasticity. For this parameter set, a power law fit with an exponent of $-1.96\pm0.03$ was found. Visual inspection of the remaining two parameter sets reveals an approximately similar exponent. For parameter set 2), finite size effects for the largest strain burst magnitudes are apparent. This is to be expected since at $\lambda=10$ $\mu$m, there exists just 80 $\lambda$ lengths within the $d=800$ $\mu$m system, whereas for parameter sets 1) and 3), where $\lambda=2$, there exist 400 $\lambda$ lengths.

The above mentioned data of fig.~\ref{strain_burst_dist1_fig}a was obtained by logarithmic binning of all discrete irreversible plastic events in the micro-plastic regime. In the recent mean-field theory (MFT) work of Dahmen and co-workers~\cite{Dahmen2009,Friedman2012} this is referred to as the stress integrated distribution, $D_{\mathrm{int}}(S)$ in which $S$ is the strain burst magnitude, whose integrand is given by the stress dependent distribution function,
\begin{equation}
D(\tau,S)\sim S^{\kappa}f\left(S(\tau_{\mathrm{c}}-\tau)^{1/\sigma}\right),
\end{equation}
where $\tau_{\mathrm{c}}$ is the critical stress at which extended plastic flow occurs, and the exponents $\kappa$ and $\sigma$, and the scaling function $f$ are universal. From MFT theory $\kappa=3/2$ and $\sigma=1/2$. Performing the stress integral gives $D_{\mathrm{int}}(S)\sim S^{-(\kappa+\sigma)}=S^{-2}$ where the last step uses the MFT exponents. The data of fig.~\ref{strain_burst_dist1_fig}a is therefore in excellent agreement with that of MFT. Further insight can be given via the stress-binned complementary cumulative distribution,
\begin{equation}
C(\tau,S)=\int_{S}^{\infty}\,dS D(\tau,S),
\end{equation}
whose stress integral gives a power law $\sim S^{(\kappa+\sigma-1)}=S^{-1}$. Fig.~\ref{strain_burst_dist1_fig}b, plots this quantity for the case of $\lambda=2$ $\mu\mathrm{m}$ and $\tau_{0}=10$ MPa and gives an exponent equal to $-0.99\pm0.02$. Fig.~\ref{strain_burst_dist1_fig}b also plots $C(\tau,S)$ for a selection of stress intervals revealing strong cut-off effects due to the tunable criticality of the MFT model. Following ref.~\cite{Friedman2012}, these curves are found to collapse when appropriately scaled via the function $f=(\tau_{\mathrm{c}}-\tau)/\tau_{\mathrm{c}}$ onto the universal mean field result, as shown in fig.~\ref{strain_burst_dist1_fig}c, giving quantitative evidence that scale free avalanche behavoir does indeed occur for the current model. Here $\tau_{\mathrm{c}}$ is taken as $\tau_{0}$. Similar behaviour is seen for the case of $\lambda=10$ $\mu$m and $\tau_{0}=10$ but with $f=(\tau_c-\tau)/\tau_{c}-c$ in which the $c=0.19$ is an adjustment parameter to correct for finite size effects~\cite{Friedman2012}. For the case of $\lambda=2$ $\mu$m and $\tau_{0}=50$ the statistics was not of a sufficient quality to perform the procedure.

When compared to more realistic two and three dimensional simulations that include dislocation reactions and much larger dislocation numbers, it might seem remarkable that such a simple model is able to exhibit scale-free behaviour. That this is the case, is a central feature of the SOC phenomenon in which the universal features of the critical dislocation configuration are robust against the details of the underlying physical model. Moreover, due to this simplicity, extremely good power law behaviour is obtained over many decades for a minimal computational effort.

\section{Discussion} \label{sec_discussion}

The simulations of sec.~\ref{sec_results} demonstrate that two dominant factors control the characteristic stress scale of the stress-strain curve. These are the choice of $\tau_{0}$ and the mobile dislocation density. Figs.~\ref{different_tau_fig} and \ref{different_rho_fig} demonstrate that $\tau_{0}$ sets an upper stress limit for extended plastic flow to occur. This upper limit is approached when the mobile dislocation density is sufficiently low that the sinusoidal stress field is the dominant contribution to the stress field each dislocation feels. By increasing the mobile dislocation content, the characteristic shear-stress scale reduces due to the increasing role of the internal stress fields arising from the elastic interaction between dislocations. At the same time a broadening of the micro-plastic regime is observed, which very much is a general property of micro-yielding, where a greater number of mobile edge dislocations correlates with larger measurable plastic strain~\cite{Abel1973}. Figs.~\ref{different_lambda_fig} to \ref{first_burst_fig} demonstrate the effect of $\lambda$ (and $d$) on the stress-strain curve is somewhat subtler. The choice of $\lambda$ influences the initial configuration of the mobile dislocation population, thereby affecting the statistics of the stress at which the first strain-burst occurs and also the way in which the extended plastic flow regime is reached. Fig.~\ref{strain_burst_dist1_fig} shows, however, that the statistics of the strain burst magnitude is insensitive to all three discussed parameters, reflecting the universality of SOC with respect to micro structural details.

How shall one quantitatively interpret the imposed sinusoidal stress field? In a real material that is nominally free of dislocation structures, some type of length scale will naturally emerge as a function of macroscopic plastic strain due to the evolution of a growing and interacting dislocation structure that eventually leads to the phenomenon of patterning~\cite{ArgonBook}. Although patterning is a term predominantly referring to the effects of latter stage II and III hardening regimes, the emergence of micro-structure length scales is expected to occur at all stages of plasticity ranging from slip, dipole and eventually cellular patterns~\cite{KuhlmannWilsdorf2002,Mughrabi1983,Kubin1992}. In fact, a micro-structural length-scale can equally well be defined for an undeformed as-grown material, where the mean dislocation spacing can be used to describe the initially present internal stress fluctuations --- a view point that is central to the early work of Tinder and co-workers~\cite{Tinder1964,Tinder1973}. From this perspective the imposed sinusoidal stress field, can be viewed as the simplest realisation of the internal stress field arising from such a structure. 

In the present model this stress field is time independent, implying it is constructed by that part of the dislocation population that is immobile, with the explicit dislocations and their dynamics, arising from the (much smaller) mobile component of the dislocation population. Thus a typical loading simulation can be seen as the deformation of a model material that has a particular sample preparation or deformation history characterised by $\tau_{0}$ and $\lambda$, and a mobile dislocation density that is only a small part of the total dislocation density.

Much past work exists concerning the emergence of internal stress and length scales as a function of deformation history. In early work on the theory of cell formation, two relationships have emerged in which the total evolving dislocation density, $\rho_{\mathrm{total}}$, plays a central role. They are:
\begin{equation}
\frac{\tau_{\mathrm{material}}}{G}\propto b\sqrt{\rho_{\mathrm{total}}} \label{EqCell1}
\end{equation}
and
\begin{equation}
\lambda_{\mathrm{material}}\propto\frac{1}{\sqrt{\rho_{\mathrm{total}}}}, \label{EqCell2}
\end{equation}
where $G$ is a representative (not necessarily pure) shear modulus and $b$ the Burgers vector magnitude. In the above, $\tau_{\mathrm{material}}$ is the evolving flow stress of the material and $\lambda_{\mathrm{material}}$ is an evolving internal length scale that can be referred to as a cell size. The first expression has its earliest origins in a Taylor hardening picture in which the total dislocation density is seen as an immobile forest dislocation population. The second equation has its theoretical origins in the early work of Holt~\cite{Holt1970} who derived it for a dipolar population of screw-dislocations, showing that a uniform arrangement was unstable to fluctuations with one such length scale dominating, characterised by eqn.~\ref{EqCell2}. This length scale, which could be related to a fixed self-screening distance of the dislocation network, was postulated to reflect an emerging cell size. The approach was based on an energy minimisation principle, however due to dislocation reactions, the more modern viewpoint is that the dynamics of cell formation lies in a statistical process involving dislocation reactions and that the screening length, and therefore cell-size, is an evolving variable~\cite{Kubin1992}.

Thus, the model parameter $\lambda$ has a direct counterpart in cell formation theory, $\lambda_{\mathrm{material}}$, which can represent quite generally, a mean-free path for a mobile dislocation, a dipolar screening length or a well evolved cell length scale. Moreover, since an unloading/loading cycle will generally return a system to the flow stress before unloading, and that the present simulations have shown that the flow stress is partly controlled by $\tau_{0}$, $\tau_{0}$ should be in some way related to $\tau_{\mathrm{material}}$. From this perspective, $\tau_{0}$ and $\lambda$ are parameters that are not entirely independent from each other. In fact, eqns.~\ref{EqCell1} and \ref{EqCell2} express that the cell size decreases inversely as a function of flow stress, a well known experimental observation that is refered to as ``similitude''~\cite{KuhlmannWilsdorf2002}. 

Although similitudity is generally confirmed by experiment, some experimental work does present a more complicated picture. Early tensile/TEM work on tapered Cu single crystals finds an initially broad distribution of cell sizes that narrows and shifts to small lengths with increasing flow stress~\cite{Prinz1982}. This result suggests that a single structural length-scale might not always be a good statistical description of the evolving micro-structure. Indeed, more modern viewpoints, in which dislocation structure evolution is a non-equilibrium process ~\cite{Haehner1996}, tend to suggest a distribution of emerging length-scales leading to a scale-free fractal-like structure. Although such micro-structures have been quantitatively established by TEM investigations of latter-stage hardened single crystals of Cu ~\cite{Haehner1998,Zaiser1999}, their existence is not universal, depending strongly on material type and deformation history. The current work does not address this aspect. More general forms of an inhomogeneous internal stress field that capture such scale-free micro-structural features can be envisioned; a direction which will be investigated in future work.

Whilst the dipolar mat geometry in an external field offers a platform with which to study the depinning transition and more generally the transition to extended plastic flow, when comparing to experiment, careful consideration has to be given to its regime of applicability. To do this, a typical simulation of secs.~\ref{sec_results} and \ref{sec_statistics} is now broadly summarised. Upon choosing numerical values for all model parameters, the $N$ dislocations are introduced to the system via a distribution of random positions. This unstable configuration is then relaxed to a local minimum energy in which the forces on each dislocation are below a small threshold value. The deformation simulation is then begun using one of the three loading modes of sec.~\ref{sec_loading}. As the stress increases, intermittent plasticity increasingly occurs until a stress is reached at which extended and overlapping strain events occur, which in the previous sections has loosely been referred to as the plastic flow regime. 

It is important to emphasise that no attempt has been made to obtain the global energy minimum of the starting configuration. Such an initial state turns out to play a crucial role in the observed properties of the model, since many high energy configurations will exist, and it is these that dominate the early stages of plasticity. As a deformation simulation proceeds, such high energy configurations structurally transform eventually leading to a plastic flow regime and often to the homogenisation of the dislocation configuration. In other words, the extended plastic flow regime should be considered to be outside the applicability regime of the present model when a comparison to experiment is made, or equivalently, the present model is only suitable for the study of the micro plastic regime of the stress-strain curve.

The rational behind the use of an initial high-energy dislocation configuration originates from the assumption that the explicit dislocation population of the model represents only the mobile dislocation network, which constitutes only a small part of the true population. Thus, in the same way as $\tau_{0}$ and $\lambda$ characterise the sample preparation or deformation history of the model material, so does the initial high energy (explicit) mobile dislocation content. This is quite compatible from the perspective of SOC in which the dislocation structure reaches a critical configuration that is far from equilibrium, and that structural rearrangements correspond to the system transforming from one SOC state to another. By construction, that part mediating the structural transformation will be the current mobile dislocation content. The central simplification of the present model, is that it separates the mobile and immobile populations, associating the former to an explicit mobile dislocation content that represents the non-equilibrium component of the network, and relegating the latter to an effective static internal stress field. That this internal stress field is unchanging and that the same explicit mobile dislocation population exists as a function of strain for the entire deformation simulation, is of course different from a real material, where the structure evolves with strain, and at any particular non-negligible strain interval, quite different dislocations might constitute the mobile dislocation population. This again emphasises that the present model should only be applied to the micro plastic regime, where significant structural evolution is minimal. 

Experimental evidence for a lack of structural evolution in the mico-plastic regime is best seen in low amplitude cyclic deformation experiments of FCC metals, in which the plastic strain per cycle can be as low as $\simeq10^{-5}$ leading to significant changes in load stress and internal length scale only after the occurrence of several tens-to-hundreds of thousands of cycles~\cite{Mughrabi1978,Basinski1992}. It is further noted, that documented experimental studies on the micro-plasticity at room temperature primarily report on movements of edge or non-screw type dislocations, whereas a clear increase in dislocation density or the formation of dislocation structures as a result of multiplication remains absent~\cite{Young1961,Vellaikal1969,Koppenaal1963}. In the bulk case there are exceptions to this trend where in the case of a work-hardened Al-Mg alloy which exhibits dynamic strain ageing, emerging structural length-scales were already detected in the micro-plastic regime~\cite{Mudrock2011} using high-resolution extensometry methods~\cite{Fressengeas2009,Roth2012,Lebyodkin2012}.

The results of sec.~\ref{sec_statistics_dist} demonstrate that the developed model exhibits power-law behaviour in the distribution of strain burst magnitudes, and thus the scale-free avalanche phenomenon seen in experiments, either via the stress-strain curve of micro-compression tests~\cite{Dimiduk2006,Dimiduk2010} or via in situ acoustic emission experiments~\cite{Miguel2001,Weiss2003}, and in simulation, via two or three dimensional dislocation dynamics simulations in which the entire network is represented by an explicit dislocation population and individual dislocation reaction mechanisms are taken into account. With an exponent of approximately -1.96, the model gives a value that is somewhat higher than that seen in simple metals~\cite{Dimiduk2006} and ice~\cite{Miguel2001}, but more comparable to that seen in LiF crystals~\cite{Dimiduk2010}. That such a simple model can admit scale-free behaviour, is connected to the dependence of the intermittent plasticity on the extremal configurations of the explicit dislocation population. This was directly seen in the statistics of the first-burst shear-stress and also the distribution of strain burst magnitudes, where with a large enough increase of $\lambda$ (say from 2 $\mu$m to 10 $\mu$m) the first burst statistics changes from being dominated by extreme value statistics to that being dominated by the statistics of the most probable (fig.~\ref{first_burst_fig}b) corresponding to an increased presence of cut-off effects in the statistics of strain burst magnitudes (fig.~\ref{strain_burst_dist1_fig}a). This is a natural result of the observation that quantities that depend on extreme value statistics can exhibit power-law behaviour in their distributions, emphasising a connection with SOC that is related to only the mobile dislocation population being in a non-equilibrium state and not to the characteristics of the present simplified immobile dislocation network --- a manifestation of a scenario referred to as ``nearly critical'' or ``robust critical'' \cite{Zaiser2006}. 

\section{Concluding remarks}

A simplified two dimensional dislocation modelling framework has been introduced in which the explicit interacting dislocation population, constrained to a simple dipolar mat geometry, represents only the mobile dislocation density component of the total dislocation density, and the much larger immobile dislocation population is described by a static internal sinusoidal shear-stress field defined by an internal shear-stress amplitude and wavelength. These model parameters, along with the initial non-equilibrium explicit mobile dislocation content characterise either the deformation or sample preparation history of the model material. Because of the static nature of the internal field and the lack of dislocation-dislocation reactions, upon loading, the present model is restricted to the micro-plastic region of the stress-strain curve, and therefore to a deformation regime for a given material that involves negligible structural evolution. Despite the simplicity of the model and the restriction to the micro-plastic regime, the deformation behaviour exhibits a rich variety of properties as a function of the model parameters. In particular, intermittent plasticity is observed whose strain burst magnitude distribution exhibits scale-free avalanche behaviour.

\section{Acknowledgements}

P.M.D wishes to thank P. D. Isp\'{a}novity for useful discussions. R.M. gratefully acknowledges the financial support of the Alexander von Humboldt foundation, his host G.M. Pharr, as well as institutional support by J.R. Greer at Caltech.

\appendix

\section{Periodic Boundary Condition Treatment} \label{appendix_a}
Due to the long range nature of eqn.~\ref{EqDDInt}, all image contributions must be taken into account for the correct treatment of the periodic boundary conditions along the dipolar mat direction. The force per unit dislocation length on a dislocation $n$ (at position $(x_{n},y_{n})$) due to a dislocation $n'$ (at position $(x_{n'},y_{n'})$) and it's images, is given by the infinite summation:
\begin{equation}
f_{x,nn'}=\frac{Gb_{x,n} b_{x,n'}}{2\pi(1-\nu)}\sum_{k=-\infty}^{\infty}\frac{(x_{n}-(x_{n'}-kd))((x_{n}-(x_{n'}-kd))^{2}-(y_{n}-y_{n'})^{2})}{((x_{n}-(x_{n'}-kd))^{2}+(y_{n}-y_{n'})^{2})^{2}}. \label{Eq1App}
\end{equation}
Due to the periodic boundary condition being one dimensional, an analytic solution to the above summation can be found. To do this, the summation in eqn.~\ref{Eq1App} is re-written as
\begin{equation}
\lim_{k'\rightarrow\infty}\sum_{k=-k'-1}^{k'}\frac{(\Delta x+kd)((\Delta x+kd)^{2}-\Delta y^{2})}{((\Delta x+kd)^{2}+\Delta y^{2})^{2}}, \label{Eq2App}
\end{equation}
where $\Delta x=x_{n}-x_{n'}$ and $\Delta y=y_{n}-y_{n'}$, and $0<\Delta x<d/2$. 

The above summation is explicitly convergent for all values of $k'$, and may be evaluated analytically in terms of poly-gamma functions. This is achieved by first expressing the summand in eqn.~\ref{Eq2App} as an irreducible partial fraction:
\begin{eqnarray}
\frac{1}{2(\Delta x-\imath\Delta y+kd)}+\frac{1}{2(\Delta x+\imath\Delta y+kd)}+ \nonumber \\
\frac{\imath\Delta y}{2(\Delta x-\imath\Delta y+kd)^{2}}-\frac{\imath\Delta y}{2(\Delta x+\imath\Delta y+kd)^{2}}.
\end{eqnarray}
In this form, the summation to a finite $k'$ may be straightforwardly obtained via a known series expansion of poly-gamma functions~\cite{AbramowitzStegun}, giving  
\begin{equation}
\sum_{k=-k'-1}^{k'}\frac{1}{(z+k d)}=-\frac{1}{d}\left[\phi^{(0)}(-1-k'+\frac{z}{d})-\phi^{(0)}(1+k'+\frac{z}{d})\right]
\end{equation}
and
\begin{equation}
\sum_{k=-k'-1}^{k'}\frac{1}{(z+k d)^{2}}=\frac{1}{d^{2}}\left[\phi^{(1)}(-1-k'+\frac{z}{d})-\phi^{(1)}(1+k'+\frac{z}{d}),\right]
\end{equation}
where $\phi^{(m)}(z)$ is the $m$'th derivative of the di-Gamma function (the logarithmic derivative of the Gamma function~\cite{AbramowitzStegun}). Performing the substitutions and taking the leading order term as $k'\rightarrow\infty$ results in eqn.~\ref{Eq2App} evaluating to
\begin{eqnarray}
\frac{1}{2d^{2}}\left[-d\pi\left[\cot\left(\frac{\pi(\Delta x+\imath\Delta y)}{d}\right)+\cot\left(\frac{\pi(\Delta x-\imath\Delta y)}{d}\right)\right]\right. \nonumber \\
\left.-i\pi^{2}\Delta y\left[\cot^{2}\left(\frac{\pi(\Delta x+\imath\Delta y)}{d}\right)+\cot^{2}\left(\frac{\pi(\Delta x-\imath\Delta y)}{d}\right)\right]\right]. \label{Eq3App}
\end{eqnarray}
For a form that is more amenable to computation the trigonometric identity,
\begin{equation}
\cot\left(\frac{a+b}{2}\right)=\frac{\cos a+\cos b}{\sin a+\sin b},
\end{equation}
is used, to obtain the final result for eqn.~\ref{Eq2App} as
\begin{equation}
\frac{-\pi\sin\left(\frac{2\pi\Delta x}{d}\right)\left[d\left(\cos\left(\frac{2\pi\Delta x}{d}\right)-\cosh\left(\frac{2\pi\Delta y}{d}\right)\right)+2\pi\Delta y\sin\left(\frac{2\pi\Delta y}{d}\right)\right]}{d^{2}\left(\cos\left(\frac{2\pi\Delta x}{d}\right)-\cosh\left(\frac{2\pi\Delta y}{d}\right)\right)^{2}}. \label{Eq4App}
\end{equation}
Eqn.~\ref{Eq4App} with the prefactor $Gb_{x,n} b_{x,n'}/2\pi(1-\nu)$ replaces eqn.~\ref{EqDDInt} in eqn.~\ref{EqTotalForce} when used in evaluating the total force of each dislocation: eqn.~\ref{EqDDIntImage}. For dislocations on the same slip plane, $\Delta y=0$, and eqn.~\ref{Eq4App} reduces to the known result~\cite{Moretti2004},
\begin{equation}
\frac{\pi}{d}\cot\left(\frac{\pi\Delta x}{d}\right).
\end{equation}

\begin{figure}[h]
\centering
\includegraphics[clip=true,width=0.95\textwidth]{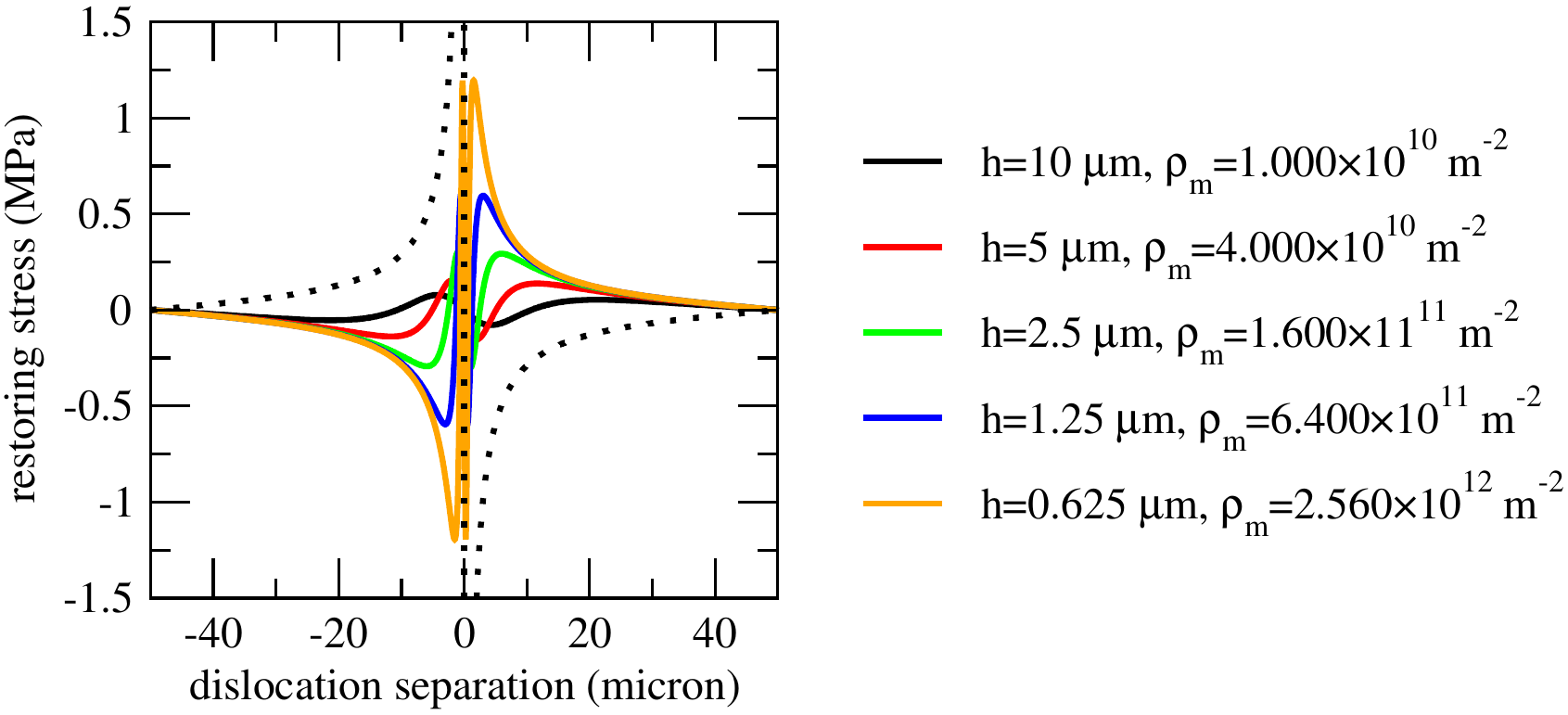}
\caption{\small Plot of the restoring shear-stress as a function of separation between two dislocations of opposite burgers vector on different slip planes for typical dislocation densities used in the described simulations (coloured lines) and two dislocations of equal Burgers vector on the same slip plane (dashed black line). For these curves, the exact image summation form, eqn.~\ref{Eq4App} is used for a system with a periodicity of $d=100$ $\mu$m.}
\label{dd_interaction_fig}
\end{figure}

Using, eqn.~\ref{Eq4App} with a dipolar mat periodicity length of $d=100$ $\mu$m, fig.~\ref{dd_interaction_fig} displays the shear-stress dependence between two dislocations of opposite Burgers vector and therefore on different slip planes (see fig.~\ref{schematic_fig}), for different values of the dislocation density and thus different $h$. Inspection of this figure reveals the short range structure to be similar to that of the bare interaction given by eqn.~\ref{EqDDInt} in sec.~\ref{sec_model}, whereas at larger distances the interaction correctly limits to zero at $\pm d/2$. In this figure the chosen numerical values of the mobile dislocation density (where $h=1/\sqrt{\rho_{\mathrm{m}}}$) will be typical of those used in the present work.  Inspection of this figure reveals that for the highest dislocation density, the restoring stress is at most of the order of 1 MPa and thus small when compared to the considered values of $\tau_{0}$, and also typical values of the stress associated with intra-slip-plane dislocation interaction at sub-micron distances (see the dashed line in fig.~\ref{dd_interaction_fig}, which displays the restoring shear-stress between two dislocations of the same Burgers vector on the same slip plane). To investigate the regime in which the two slip planes strongly interact, higher mobile dislocation densities would need to be considered or a different relationship between $h$ and the dislocation density taken.

\end{document}